\documentclass[USenglish,oneside,twocolumn]{article}

\usepackage[utf8]{inputenc}%(only for the pdftex engine)
\usepackage[big]{dgruyter_NEW}
\usepackage{blindtext}

\usepackage[colorinlistoftodos,prependcaption]{todonotes}
\usepackage{regexpatch}


\usepackage{todonotes}
\usepackage{eurosym}
\usepackage{array,makecell,booktabs}
\newcolumntype{M}[1]{>{\centering\arraybackslash}m{#1}}
\usepackage[flushleft]{threeparttable}

\usepackage{lmodern,babel,adjustbox,booktabs,multirow,subcaption}
\usepackage{listings}

\usepackage{tikz}

\usepackage{fontawesome}
 
\cclogo{\includegraphics{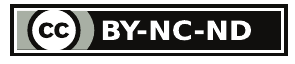}}
  
\begin{document}

  \author[1]{Yana Dimova}

 \author[2]{Gunes Acar}

 \author[3]{Lukasz Olejnik}

 \author[4]{Wouter Joosen}
 
 \author[5]{Tom Van Goethem}

 \affil[1]{imec-DistriNet, KU Leuven, E-mail: yana.dimova@cs.kuleuven.be}

 \affil[2]{imec-COSIC, KU Leuven, E-mail: gunes.acar@esat.kuleuven.be}

 \affil[3]{European Data Protection Supervisor, independent researcher, E-mail: me@lukaszolejnik.com}

 \affil[4]{imec-DistriNet, E-mail: wouter.joosen@cs.kuleuven.be}
 
 \affil[5]{imec-DistriNet, E-mail: tom.vangoethem@cs.kuleuven.be}

  \title{\huge{The CNAME of the Game:\\Large-scale Analysis of DNS-based Tracking Evasion}}

  \runningtitle{The CNAME of the Game: Large-scale Analysis of DNS-based Tracking Evasion}

  %\subtitle{...}

  \begin{abstract}
{
Online tracking is a whack-a-mole game between trackers who build and monetize behavioral user profiles through intrusive data collection, and anti-tracking mechanisms, deployed as a browser extension, built-in to the browser, or as a DNS resolver.
As a response to pervasive and opaque online tracking, more and more users adopt anti-tracking tools to preserve their privacy.
Consequently, as the information that trackers can gather on users is being curbed, some trackers are looking for ways to evade these tracking countermeasures.
In this paper we report on a large-scale longitudinal evaluation of an anti-tracking evasion scheme that leverages CNAME records to include tracker resources in a same-site context, effectively bypassing anti-tracking measures that use fixed hostname-based block lists.
Using historical HTTP Archive data we find that this tracking scheme is rapidly gaining traction, especially among high-traffic websites.
Furthermore, we report on several privacy and security issues inherent to the technical setup of CNAME-based tracking that we detected through a combination of automated and manual analyses. We find that some trackers are using the technique against the Safari browser, which is known to include strict anti-tracking configurations.
Our findings show that websites using CNAME trackers must take extra precautions to avoid leaking sensitive information to third parties.}
\end{abstract}
%  \keywords{tracking, CNAME, evasion}
%  \classification[PACS]{}
 % \communicated{...}
 % \dedication{...}

  \journalname{Proceedings on Privacy Enhancing Technologies}
%  \DOI{Editor to enter DOI}
 \startpage{1}
%  \received{..}
%  \revised{..}
%  \accepted{..}

%  \journalyear{..}
%  \journalvolume{..}
%  \journalissue{..}

\maketitle
\section{Introduction}
Websites use trackers for various purposes including analytics, advertising and marketing.
Although tracking may help websites in monetization of their content, 
the use of such methods may often come at the expense of users' privacy, for example when it involves building detailed behavioral profiles of users.
As a reaction to the omnipresence of online tracking, in the previous decade many countermeasures have been developed, including specialised  browser extensions, DNS resolvers, and built-in browser protections.
As of today, all major browsers (except Google Chrome) include some forms of anti-tracking measures. Safari's Intelligent Tracking Prevention (ITP) includes multiple features to thwart various forms of tracking and circumvention techniques~\cite{safarithirdpartycookieblock}; Firefox' Enhanced Tracking Protection (ETP) and the tracking prevention mechanism in Edge rely on block lists to exclude trackers~\cite{firefoxetp,edgetrackingprevention}.

As a counter-reaction to the increased use of anti-tracking measures, several trackers have resorted to new techniques in an attempt to circumvent these measures.
Prominent and well-studied examples of these evasion techniques include browser fingerprinting~\cite{eckersley2010unique,Acar:2014:WNF:2660267.2660347,nikiforakis2013cookieless,englehardt2016online,iqbal2020fingerprinting}, leveraging various browser mechanisms to persist a unique identifier~\cite{criteohsts, Apple-HSTS-Abuse,ayenson2011flash}, and creating a fingerprint from hardware anomalies~\cite{mavroudis2017privacy,dey2014accelprint,zhang2019sensorid}.
A notable example for the use of evasion techniques is the case of Criteo, one of the tracking actors we study in this paper. In 2015, Criteo leveraged a redirection technique to set first-party cookies~\cite{criteofirstparty,benguerah2017setting}, and later abused the HTTP Strict-Transport-Security mechanism~\cite{criteohsts, Apple-HSTS-Abuse}, both in an effort to circumvent Safari's Intelligent Tracking Protection (ITP). Our study complements these past reports with an observation that Criteo is applying a specialised form of first-party tracking to Safari browsers.

In this paper, we report on an evasion technique that has been known for several years but recently gained more traction, presumably due to the increased protection against third-party tracking.
This tracking scheme takes advantage of a CNAME record on a subdomain such that it is same-site to the including website.
As such, defenses that block third-party cookies are rendered ineffective.
Furthermore, because custom subdomains are used, these are unlikely to be included in block lists (instead of blocking the tracker for all sites, block lists would have to include every instance for each website including the CNAME-based tracker).

Using the HTTP Archive dataset, supplemented with results from custom crawls, we report on a large-scale evaluation of the CNAME-based tracking ecosystem, involving 13 manually-vetted tracking companies.
We find that this type of tracking is predominantly present on popular websites: 9.98\% of the top 10,000 websites employ at least one CNAME-based tracker.

The use of such tracking is rising. Through a historical analysis of the ecosystem, we show that the number of websites that rely on this type of tracking is steadily growing, especially compared to similarly-sized tracking companies which have experienced a decline in number of publishers.
We find that CNAME-based tracking is often used in conjunction with other trackers: on average 28.43 third-party tracking scripts can be found on websites that also use CNAME-based tracking.
We note that this complexity in the tracking ecosystem results in unexpected privacy leaks, as it actually introduces new privacy threats inherent to the ecosystem where various trackers often set first-party cookies via the \texttt{document.cookie} interface. We find that due to how the web architecture works, such practices lead to wide-spread cookie leaks. Using automated methods we measure such cookie leaks to CNAME-based trackers and identify cookie leaks on 95\% of the sites embedding CNAME-based trackers.
Although most of these leaks are due to first-party cookies set by other third-party scripts, we also find cases of cookie leaks to CNAME-based trackers in POST bodies and in URL parameters, which indicates a more active involvement by the CNAME-based trackers.

Furthermore, through a series of experiments, we report on the increased threat surface that is caused by including the tracker as same-site.
Specifically, we find several instances where requests are sent to the tracking domain over an insecure connection (HTTP) while the page was loaded over a secure channel (HTTPS).
This allows an attacker to alter the response and inject new cookies, or even alter the HTML code effectively launching a cross-site scripting attack against the website that includes the tracker; the same attacks would have negligible consequences if the tracking iframe was included from a cross-site domain.
Finally, we detected two vulnerabilities in the tracking functionality of CNAME-based trackers.
This could expose the data of visitors on \emph{all} publisher websites through cross-site scripting and session-fixation attacks.

\medskip
\noindent In summary, we make the following contributions:
\begin{itemize}
	\item We provide a general overview of the CNAME-based tracking scheme, based on a large-scale analysis involving a custom detection method, allowing us to discover previously unknown trackers.
	\item We perform a historical analysis to study the ecosystem, and find that this form of first-party tracking is becoming increasingly popular and is often used to complement third-party tracking.
	\item Through a series of experiments, we analyze the security and privacy implications that are intrinsic to the tracking scheme. We identify numerous issues, including the extensive leakage of cookies set by third-party trackers.
	\item Based on the observation of practical deployments of the CNAME-based tracking scheme, we report on the worrying security and privacy practices that have negative consequences for web users.
	\item We discuss the various countermeasures that have recently been developed to thwart this type of tracking, and assess to what extent these are resistant to further circumvention techniques.
\end{itemize}

\section{Background}

\subsection{Web browser requests}
Upon visiting a web page, the browser will make various requests to fetch embedded resources such as scripts, style sheets and images.
Depending on the relation between the embedding website and the site that the resources are hosted on, these can be \textit{same-origin}, \textit{same-site} or \textit{cross-site}.
If the resource shares the same scheme (i.e.\ http or https), host (e.g.\ www.example.com) and port (e.g.\ 80 or 443) as the embedding site, it is considered same-origin.
In case there is no exact match for the host, but the resource is located on the same registrable domain name, the effective top level domain plus one (\textit{eTLD+1}), as the embedding website (e.g.\ www.example.com and foo.example.com), it is considered same-site.
Finally, resources that have a different eTLD+1 domain with regard to the including website are considered cross-site, i.e., resources from \textit{tracker.com} included on \textit{example.com} are cross-site.

Prior to making the connection to the server, the domain name first needs to be resolved to an IP address.
In the most straightforward case, the DNS resolution of the domain name returns an \texttt{A} record containing the IP address.
However, the domain could also use a \texttt{CNAME} record to refer to any other domain name.
This can be an iterative process as the new domain name can again resolve to another \texttt{CNAME} record; this process continues until an \texttt{A} record is found.
Through this indirection of CNAMEs, the host that the browser connects to may belong to a different party, such as a tracker, than the domain it actually requests the resource from.
This means that requests to \textit{xxx.example.com} may actually be routed to a different site, such as \textit{yyy.tracker.com}.

\textbf{Cookie scoping}
Before a request is sent, the browser will first determine which cookies to attach in the HTTP request. 
This includes all cookies that were set on the same (sub)domain as the one where the request will be sent to.
Other cookies that will be included are those that were set by a same-site resource, i.e.\ either on another subdomain, or on the top domain, and had the \texttt{Domain} attribute set to the top domain, for instance by the following response header on https://sub.example.com/
\begin{lstlisting}[basicstyle=\ttfamily\footnotesize,]
Set-Cookie: cookie=value; Domain=example.com
\end{lstlisting}
Cookies that were set without the \texttt{Domain} attribute will only be included on requests that are same-origin to the response containing the \texttt{Set-Cookie} header.
The \texttt{SameSite} attribute on cookies determines whether a cookie will be included if the request is cross-site.
If the value of this attribute is set to \texttt{None}, no restrictions will be imposed; if it is set to \texttt{Lax} or \texttt{Strict}, it will not be included on requests to resources that are cross-site to the embedding website; the latter imposes further restrictions on top-level navigational requests.
Several browser vendors intend to move to a configuration that assigns \texttt{SameSite=Lax} to all cookies by default~\cite{chromiumsamesite,draft-west-cookie-incrementalism-01,firefoxsamesitelaxdefault}.
As such, for third-party tracking to continue to work, the cookies set by the trackers explicitly need to set the \texttt{SameSite=None} attribute, which may make them easier to distinguish.
For CNAME-based tracking, where the tracking requests are same-site, the move to SameSite cookie by default has no effect.

\subsection{Tracking}
\subsubsection{Third-party tracking}
In a typical tracking scenario, websites include resources from a third-party tracker in a cross-site context.
As a result, when a user visits one of these web pages, a cookie originating from the third party is stored in the visitor's browser.
The next time a user visits a website on which the same tracker is embedded, the browser will include the cookie in the request to the tracker.
This scheme allows trackers to identify users across different websites to build detailed profiles of their browsing behavior.
Such tracking has triggered privacy concerns and has resulted in substantial research effort to understand the complexity of the tracking ecosystem \cite{mayer2012third,englehardt2015cookies} and its evolution \cite{lerner2016internet}.

\subsubsection{First-party tracking}
In first-party tracking the script and associated analytics requests are loaded from a same-site origin.
Consequently, any cookie that is set will only be included with requests to the same site.
Historically, one method that was used to bypass this limitation was cookie matching~\cite{olejnik2014selling}, where requests containing the current cookie are sent to a common third-party domain.
However, such scripts can be blocked by anti-tracking tools based on simple matching rules.
Instead, the technique covered in this work uses a delegation of the domain name, which circumvents the majority of anti-tracking mechanisms currently offered to users.

\subsubsection{CNAME-based tracking}

\begin{figure}
	\centering
	\includegraphics[width=\linewidth]{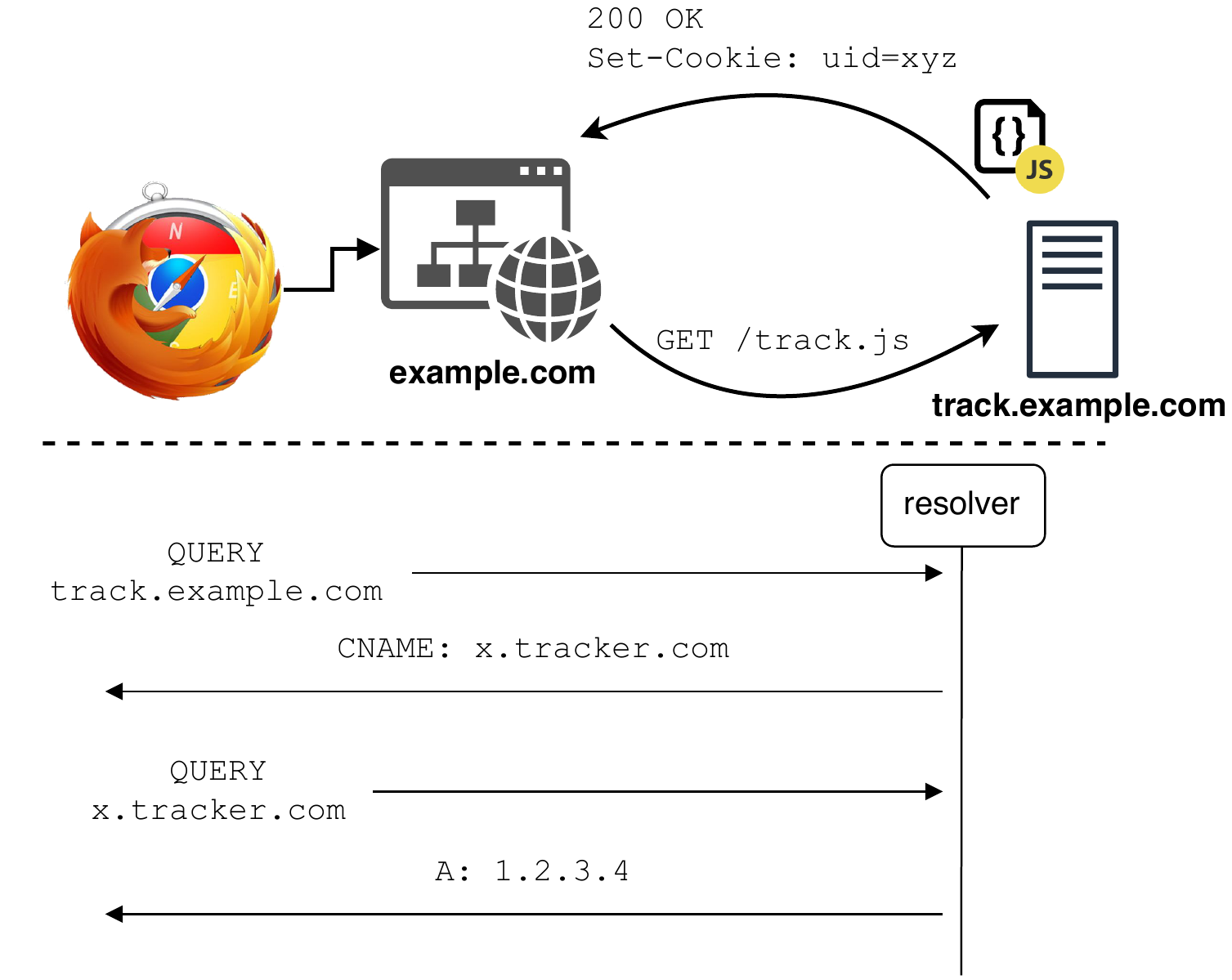} 
	\caption{Overview of CNAME-based tracking.}
	\label{fig:cname-scheme-overview}
\end{figure}

\textbf{General overview}
In the typical case of third-party tracking, a website will include a JavaScript file from the tracker, which will then report the analytics information by sending (cross-site) requests to the tracker domain.
With CNAME-based tracking, the same operations are performed, except that the domain that the scripts are included from and where the analytics data is sent to, is a subdomain of the website.
For example, the website \texttt{example.com} would include a tracking script from \texttt{track.example.com}, thus effectively appearing as same-site to the including website.
Typically, the subdomain has a CNAME record that points to a server of the tracker.
An overview of the CNAME-based tracking scheme is shown in Figure~\ref{fig:cname-scheme-overview}.

\noindent \textbf{Bypassing anti-tracking measures}
The CNAME tracking scheme has direct implications for many anti-tracking mechanisms.
Because the requests to the tracking services are same-site (i.e.\ they point to the same eTLD+1 domain as the visited website), countermeasures that aim to block third-party cookies, such as Safari's ITP, are effectively circumvented.
Other popular anti-tracking mechanisms that rely on blocking requests or cookies by using block lists (such as EasyPrivacy \cite{easyprivacy} or Disconnect.me \cite{disconnectme}) become much harder to maintain  when trackers are served from a custom subdomain that is unique to every website. 
To block CNAME-based tracking, block lists would need to contain an entry for every website that uses the CNAME-based tracking service, instead of a single entry per tracker or match all DNS-level domains, leading to greater performance costs.

As a consequence of how the CNAME-based tracking scheme is constructed, it faces certain limitations in comparison to third-party tracking.
For instance, there no longer exists a common identifier shared across the different websites (in typical third-party tracking, the third-party cookie is responsible for this functionality).
Consequently, visits to different websites cannot be attributed to the same user using standard web development features.

\section{Detecting CNAME-based tracking}
\label{sec:cname-tracking}

In this section we describe the composition of the datasets along with the various steps of our methodology that we used to detect CNAME-based trackers and the publishers that include them.

\subsection{Dataset}
In order to analyze the CNAME-based tracking scheme at a scale, we leveraged the (freely available) crawling data from HTTP Archive \cite{http_archive}.
This dataset originates from visiting the home page of all origins from the Chrome User Experience Report (CrUX), which lists websites (including those hosted on subdomains) frequently visited by Chrome users.
The results reported in this section are based on the desktop crawl performed in October, consisting of 5,506,818 visited web pages from 4,218,763 unique eTLD+1 domains.
The information contained in this dataset includes all request and response headers of all the requests (507M in total) that were made when visiting the web pages with the latest Chrome browser.
As the dataset only contains the IP address of the remote host that was connected to at the time of making the request, we extended the dataset with DNS records (in particular CNAME) obtained by running zdns \cite{zdns} on all first-party subdomains.

\subsubsection{Methodology}
\label{sec:detection-methodology}
\noindent \textbf{Discovering trackers}
To detect services that offer CNAME-based tracking, we used a three-pronged approach that leverages features intrinsic to the ecosystem, combining both automated and manual analysis.
First we filtered all requests from HTTP Archive's dataset and only considered the ones that were same-site but not same-origin, i.e.\ the same eTLD+1 but not the exact same origin as the visited web page.
Furthermore, we only retained requests to domain names that returned a CNAME record referring (either directly or indirectly after redirection of other CNAME records) to a different eTLD+1 domain in our DNS data.
We aggregated these requests on the eTLD+1 of the CNAME record, and recorded a variety of information, such as the average number of requests per website, variation of request size, percentage of requests that contain a cookie or set one via the HTTP response header, etc.
In Appendix~\ref{sec:assisted-detection} we elaborate on these features and discuss how they could be used to assist or automate the detection of CNAME-based tracking.
Out of the resulting 46,767 domains, we only consider the ones that are part of a CNAME-chain on at least 100 different websites, which leaves us with 120 potential CNAME-based trackers. 

In the second phase, we performed a manual analysis to rule out services that have no strict intention to track users. 
Many services that are unrelated to tracking, such as CDNs, use a same-site subdomain to serve content, and may also set a cookie on this domain, thus giving them potential tracking \emph{capabilities}.
For instance, Cloudflare sets a \texttt{\_cfduid} cookie in order to detect malicious visitors, but does not intend to track users with this cookie (user information is kept less than 24 hours)~\cite{cloudflarecookies}.
For each of the 120 domains, we visited the web page of the related organization (if available) and gathered information about the kind of service(s) it provides according to the information and documentation provided on its website.
Based on this information, we then determined whether tracking was the main service provided by this company, either because it explicitly indicated this, or tracking would be required for the main advertised product, e.g.\ in order to provide users with personalized content, or whether this was clear from the way the products were marketed.
For instance one such provider, Pardot offers a service named ``Marketing Automation'', which they define as ``a technology that helps businesses grow by automating marketing processes, tracking customer engagement, and delivering personalized experiences to each customer across marketing, sales, and service''\footnote{\url{https://www.pardot.com/what-is-marketing-automation/}}, indicating that customers (website visitors) may be tracked.
Finally, we validate this based on the requests sent to the purported tracker when visiting a publisher website: we only consider a company to be a tracker when a uniquely identifying parameter is stored in the browser and sent along with subsequent requests, e.g.\ via a cookie or using localStorage.
Using this method, we found a total of 5 trackers. 
Furthermore, we extended the list with eight trackers from the CNAME cloaking blocklist by NextDNS~\cite{nextdnscnamecloakinggithub,nextdnscnamecloaking}. Four of the trackers we detected in our manual analysis were not included in the block list. We left two of the trackers from the list out of consideration, as they were not included in the DNS data.
In total we consider 13 CNAME-based trackers.

\noindent \textbf{Detecting the prevalence of CNAME-based tracking}
By examining request information to hostnames having a CNAME record to one of the identified trackers, we manually constructed a \emph{signature} for all tracking requests for each of the 13 trackers, based on the DNS records and request/response information (e.g.\ the same JavaScript resource being accessed or a request URL according to a specific pattern).
This allows us to filter out any instances where a resource was included from a tracking provider but is unrelated to tracking, as the providers may offer various other services and simply relying on DNS data to detect CNAME publisher domains leads to an overestimate (we justify this claim in Section~\ref{sec:method_validation}).
Using this approach, we detected a total of 10,474 websites (eTLD+1) that used at least one of the trackers; we explore these publishers that use CNAME tracking in more detail in Section~\ref{sec:tracking-customers}.

\subsection{Alternative user agent}
A limitation of the HTTP Archive dataset, is that all websites were visited with the Chrome User-Agent string, a browser that does not have built-in tracking protection. 
Furthermore, only the home page of each website was visited.
To evaluate whether these limitations would affect our results, we performed a crawling experiment on the Tranco top 10,000 websites\footnote{\url{https://tranco-list.eu/list/6WGX/10000}}; for every website, we visited up to 20 web pages (totaling 146,397 page visits). 
We performed the experiment twice: once with the Chrome User-Agent string, and once with Safari's.
The latter is known for its strict policies towards tracking, and thus may receive different treatment.
We used a headless Chrome instrumented through the Chrome DevTools Protocol \cite{chromedevtools} as our crawler. 
A comparative analysis of these two crawls showed that one tracker, namely Criteo, would only resort to first-party tracking for Safari users.
Previously, this tracker was found to abuse top-level redirections~\cite{criteofirstparty} and leverage the HTTP Strict Transport Security (HSTS) mechanism to circumvent Safari's ITP~\cite{criteohsts,Apple-HSTS-Abuse}.

\subsection{Coverage}
Finally, to analyze the representativeness of our results and determine whether the composition of the HTTP Archive dataset did not affect our detection, we performed a comparative analysis with our custom crawl.
In the 8,499 websites that were both in the Tranco top 10k, and the HTTP Archive dataset, we found a total of 465 (5.47\%) websites containing a CNAME-based tracker.
These included 66 websites that were not detected to contain CNAME-based tracking based on the data from HTTP Archive (as it does not crawl through different pages).
On the other hand, in the HTTP Archive dataset we found 209 websites that were detected to contain a CNAME-based tracker, which could not be detected as such based on our crawl results.
This is because the HTTP Archive dataset also contains popular subdomains, which are not included in the Tranco list.
As such, we believe that the HTTP Archive dataset provides a representative view of the state of CNAME-based tracking on the web.
We note however that the numbers reported in this paper should be considered lower bounds, as certain instances of tracking can only be detected when crawling through multiple pages on a website.

\section{CNAME-based tracking}
\label{sec:cname-tracking}
In this section, we provide an in-depth overview of the CNAME-based tracking ecosystem through a large-scale analysis.

\subsection{CNAME-based trackers}
An overview of the detected trackers can be found in Table~\ref{tab:trackers-list}.
For every tracker we indicated the number of publishers, counted as the number of unique eTLD+1 domains that have at least one subdomain set up to refer to a tracker (typically with a CNAME record).
Furthermore, we estimated the total number of publishers by levering DNS information from the SecurityTrails API \cite{security_trails}.
More precisely, all CNAME-based trackers either require the publishers that include them to set a CNAME record to a specific domain, or the trackers create a new subdomain for every publisher.
As such, the estimated number of publishers could be determined by finding the domains that had a CNAME record pointing to the tracker, or by listing the subdomains of the tracker domain and filtering out those that did not match the pattern that was used for publishers.
For Ingenious Technologies we were unable to estimate the total number of publishers as they use a wildcard subdomain (and thus it could not be determined whether a subdomain referred to an actual publisher using CNAME tracking).

\renewcommand{\arraystretch}{1.1}
\begin{table*}
	\caption{Overview of the analyzed CNAME-based trackers, based on the HTTP Archive dataset from October 2020.}
	\centering
	\label{tab:trackers-list}
	\begin{threeparttable}
		\adjustbox{max width=\textwidth}{
			\begin{tabular}{ M{3.8cm} r r r c c c } 
				\hline
				& & & & \multicolumn{3}{c}{\textbf{requests to tracker is blocked by}} \\
				\textbf{Tracker} & \multicolumn{1}{M{2.2cm}}{\textbf{Detected \#~publishers}} & \multicolumn{1}{M{2.2cm}}{\textbf{Est.~total \#~publishers}} & \multicolumn{1}{M{1.6cm}}{\textbf{Pricing} (min.~/mo)} & \multicolumn{1}{M{2.3cm}}{\textbf{uBlock Origin Firefox}} & \multicolumn{1}{M{2.3cm}}{\textbf{uBlock Origin Chrome}} & \multicolumn{1}{M{2.9cm}}{\textbf{NextDNS CNAME~blocklist}} \\
				\hline
				Pardot & 5,993 & 21,759 & \$1,250\hphantom{\textsuperscript{$\dagger$}} & \hphantom{\textsuperscript{*}}\faCheck\textsuperscript{*} & \hphantom{\textsuperscript{*}}\faCheck\textsuperscript{*} & \faClose \\
				Adobe Experience Cloud & 2,612 & 9,029 & \$5,000\textsuperscript{$\dagger$} & \faCheck & \faCheck & \faCheck\\ 
				Act-On Software & 1,041 & 2,533 & \$900\hphantom{\textsuperscript{$\dagger$}} & \faCheck & \faCheck & \faClose \\ 
				Oracle Eloqua & 304 & 3,743 & \$2,000\textsuperscript{$\dagger$} & \faCheck & \faClose & \faClose \\ 
				Eulerian & 253 & 1,501 & ?\hphantom{\textsuperscript{$\dagger$}} & \faCheck & \faClose & \faCheck\\
				Webtrekk & 101 & 822 & ?\hphantom{\textsuperscript{$\dagger$}} & \faCheck & \faCheck & \faCheck \\
				Ingenious Technologies & 41 & - & ?\hphantom{\textsuperscript{$\dagger$}} & \faClose & \faClose & \faCheck \\ 
				TraceDock & 49 & 69 & \euro49\hphantom{\textsuperscript{$\dagger$}} & \faClose & \faClose & \faCheck \\ 
				<intent> & 14 & 124 & ?\hphantom{\textsuperscript{$\dagger$}} & \faClose & \faClose & \faCheck \\
				AT Internet & 31 & 74 & \euro355\hphantom{\textsuperscript{$\dagger$}} & \faClose & \faClose & \faCheck\\
				Criteo & 16 & 13,082 & ?\hphantom{\textsuperscript{$\dagger$}} & \faCheck & \faClose & \faCheck \\
				Keyade & 12 & 86 & ?\hphantom{\textsuperscript{$\dagger$}} & \faCheck & \faClose & \faCheck \\
				Wizaly & 12 & 55 & \$2000\textsuperscript{$\dagger$} & \faClose & \faClose & \faCheck \\
				\hline
		\end{tabular}}
	\end{threeparttable}
	\begin{tablenotes}
		\small
		\item $\dagger$: Pricing information does not originate from original source, but as reported in reviews of the product. 
		\item *: Requests made to the CNAME subdomain triggered by a third-party analytics script hosted on \texttt{pardot.com}; the block-\\list prevents the analytics script from loading. If this script was loaded from the CNAME domain, it would not be blocked.
	\end{tablenotes}
\end{table*}

We noted the price of the services offered by the tracker suppliers when such information was available, either from the tracker's website or through third-party reviews.
In most cases, with the exception of TraceDock, which specifically focuses on providing mechanisms for circumvention of anti-tracking techniques, the offered services included a range of analytics and marketing tools.

Finally, for every tracker we determined whether tracking requests would be blocked by three relevant anti-tracking solutions: uBlock Origin (version 1.26) on both Firefox and Chrome, and the NextDNS CNAME blocklist \cite{cname_cloaking_blocklist}, which was used to extend the list of trackers we considered.
As of version 1.25 of uBlock Origin, the extension on Firefox implements a custom defense against CNAME-based tracking~\cite{ublock-origin-blocks-cname}, by resolving the domain name of requests that are originally not filtered by the standard block list and then again checks this block list against the resolved CNAME records.
Because Chrome does not support a DNS resolution API for extensions, the defense could not be applied to this browser.
Consequently, we find that four of the CNAME-based trackers (Oracle Eloqua, Eulerian, Criteo, and Keyade) are blocked by uBlock Origin on Firefox but not on the Chrome version of the anti-tracking extension.

\subsection{Tracking publishers}

\label{sec:tracking-customers}
As a result of our analysis of the HTTP Archive dataset, we detected 10,474 eTLD+1 domains that had a subdomain pointing to at least one CNAME-based tracker, with 85 publishers referring to two different trackers.
We find that for 9,501 publisher eTLD+1s the tracking request is included from a same-site origin , i.e., the publisher website has the same eTLD+1 as the subdomain it includes tracker content from.
Furthermore, on 18,451 publisher eTLD+1s we found the tracker was included from a cross-site origin; these were typically sites that were related in some way, e.g.\ belonging to the same organization.
Although these instances cannot circumvent countermeasures where all third-party cookies are blocked, e.g.\ the built-in protection of Safari, they still defeat blocklists.
\begin{figure}
	\centering
	\includegraphics[width=0.8\linewidth]{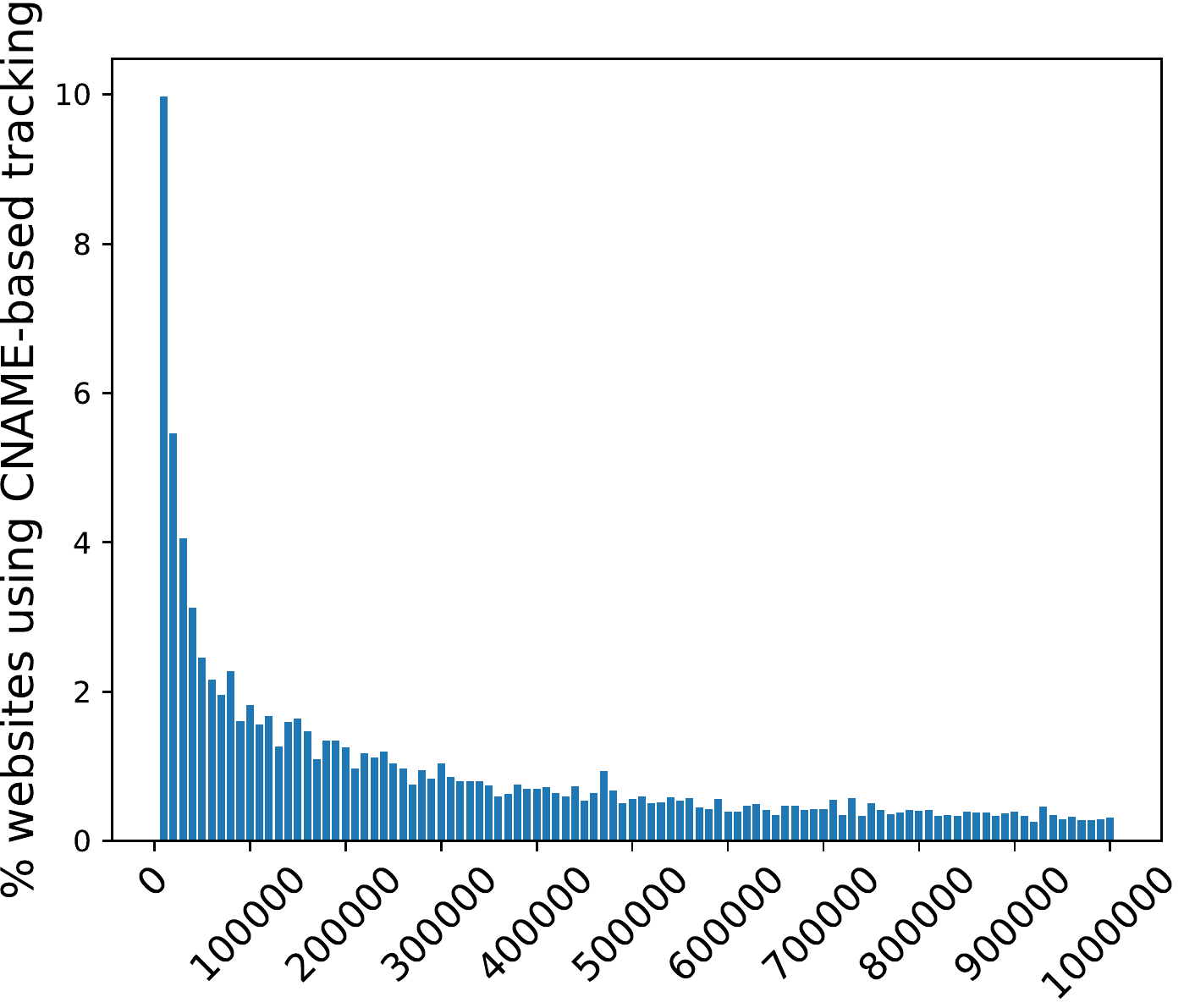}
	\caption{Percentage of websites using CNAME-based tracking per bin of 10,000 ranks.}
	\label{fig:customer-ranking}
\end{figure}

Figure~\ref{fig:customer-ranking} displays the percentage of publisher eTLD+1s involved in CNAME-based tracking, both in a same-site or cross-site context, for bins of 10,000 Tranco-ranked websites.
The ratio of same-site to cross-site CNAME-based tracking is consistently between 50\% and 65\% for all bins. 
We can clearly see that the use of CNAME-based tracking is heavily biased towards more popular websites.
In the top 10,000 Tranco websites 10\% refer to a tracker via a CNAME record.
Because our dataset only contains information about the homepage of websites, and does not include results from Criteo, the reported number should be considered a lower bound.

\begin{figure}
	\centering
	\includegraphics[width=0.8\linewidth]{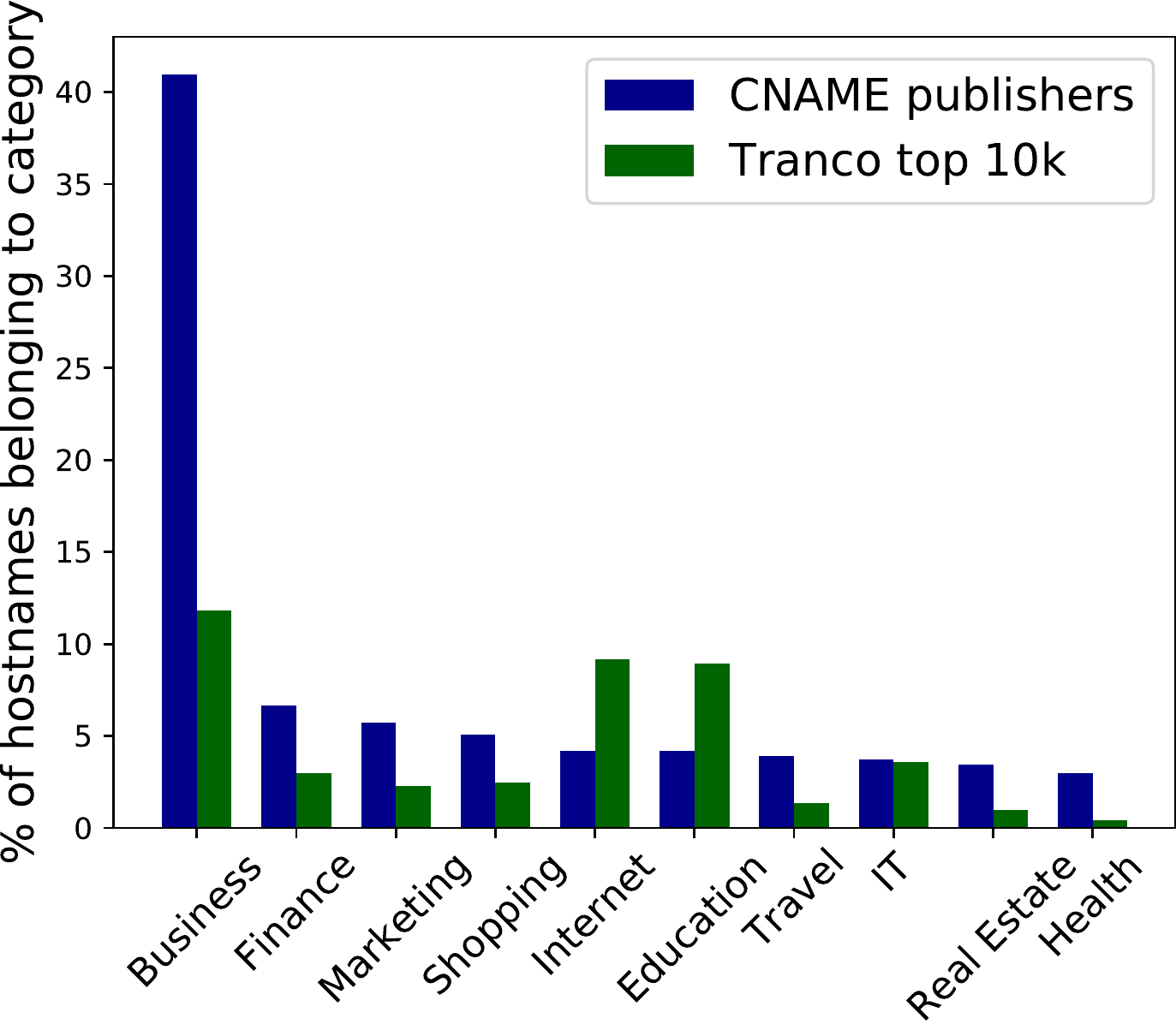}
	\caption{Most popular categories among CNAME-based tracking publishers.}
	\label{fig:customer-categories}
\end{figure}

Using the categorization service by McAfee~\cite{mcafeecategory}, we determined the most popular categories among CNAME-based tracking publishers, as shown in Figure~\ref{fig:customer-categories}.
As a baseline comparison, we also include the distribution of categories in the Tranco top 10k.
Because of the strong financial motives to perform tracking, e.g.\ marketing and attribution of online purchases, it is not surprising that publishers are mainly financially-focused, with approximately 40\% of the publisher's websites being categorized as Business.

Finally, we explored to what extent publishers that employ CNAME-based tracking also include third-party trackers.
To this end we analyzed all requests using the EasyPrivacy blocklist \cite{easyprivacy} to determine the number of trackers that would be blocked by this list.
We find that on the vast majority of websites that include a CNAME-based tracker (93.97\%) at least one third-party tracker was present; on average these sites had 28.43 third-party tracking requests.
This clearly shows that CNAME-based tracking is most often used in conjunction with other types of tracking.
From a privacy perspective this may cause certain issues, as the other trackers may also set first-party cookies via JavaScript; we explore this in more detail in Section~\ref{sec:implications-of-cname-tracking}.

\section{Historical Evolution}
\label{sec:historical-evolution}

In this section we report on various analyses we performed to capture the longitudinal evolution of CNAME-based tracking.

\subsection{Uptake in CNAME-based tracking}
First, we explore the change in prevalence of CNAME-based tracking over time.
To achieve this, we leverage the dataset of HTTP Archive, which is collected on a monthly basis and dates back several years.
We consider the datasets from December 2018, when the pages from the Chrome User Experience Report started to be used as input for their crawler, until October 2020.

\begin{figure}
	\centering
	\includegraphics[width=\linewidth]{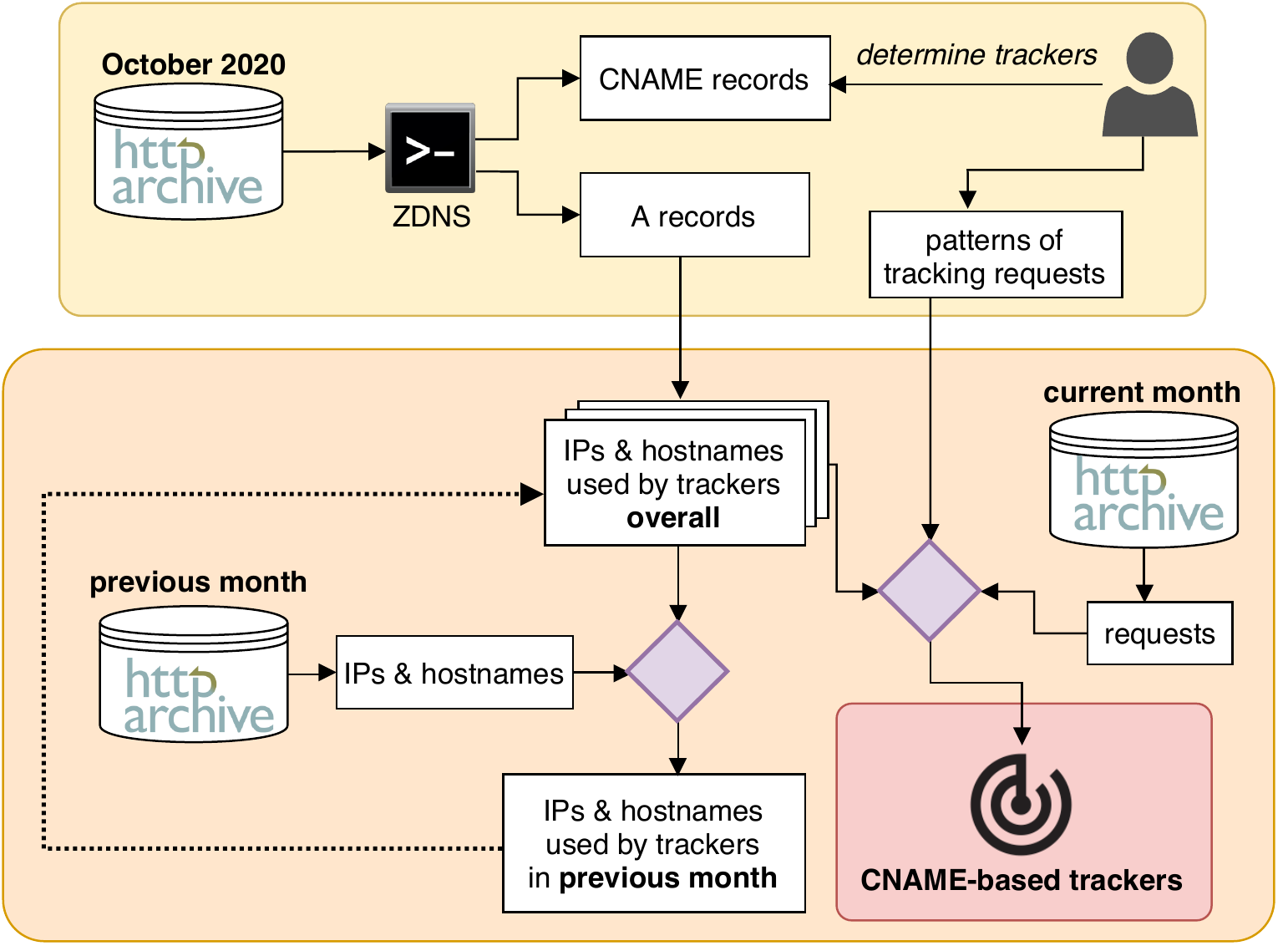} 
	\caption{Overview of the methodology that was used to determine CNAME-based trackers over time.}
	\label{fig:methodology-overview}
\end{figure}

To determine the number of publishers using CNAME tracking over time, we used an iterative approach as shown in Figure~\ref{fig:methodology-overview}.
Starting from the most recent month (October 2020), we obtained the domain names and associated IP addresses that were used to connect to the CNAME-trackers.
Next, we use data from HTTP Archive's dataset from the previous month to determine all IP addresses that (confirmed) CNAME domains resolve to, allowing us to capture changes of IP addresses by trackers.
By adding these IP addresses to the list of IPs we found in October through a scan with zdns, we obtain a set of IP addresses that were ever used by the different CNAME trackers.
Furthermore, whenever we noticed that a tracker is using IPs within a certain range for the tracking subdomains, we added the whole range to the set of used IPs (e.g. Eulerian allocates IP addresses in the range 109.232.192.0/21 for the tracking subdomains). 
Relying just on the IP information would likely lead to false positives as the trackers provide various other services which may be hosted on the same IP address, and ownership of IP addresses may change over time.
To prevent marking unrelated services as tracking, we rely on our manually-defined \emph{request signatures} (as defined in Section~\ref{sec:detection-methodology}) to filter out any requests that are unrelated to tracking.
Using the domain names of the confirmed tracking requests and the set of IP addresses associated with tracking providers, we can apply the same approach again for the previous month.
We repeat this process for every month between October 2020 and December 2018.

\begin{figure}[t]
	\centering
	\includegraphics[width=0.9\linewidth]{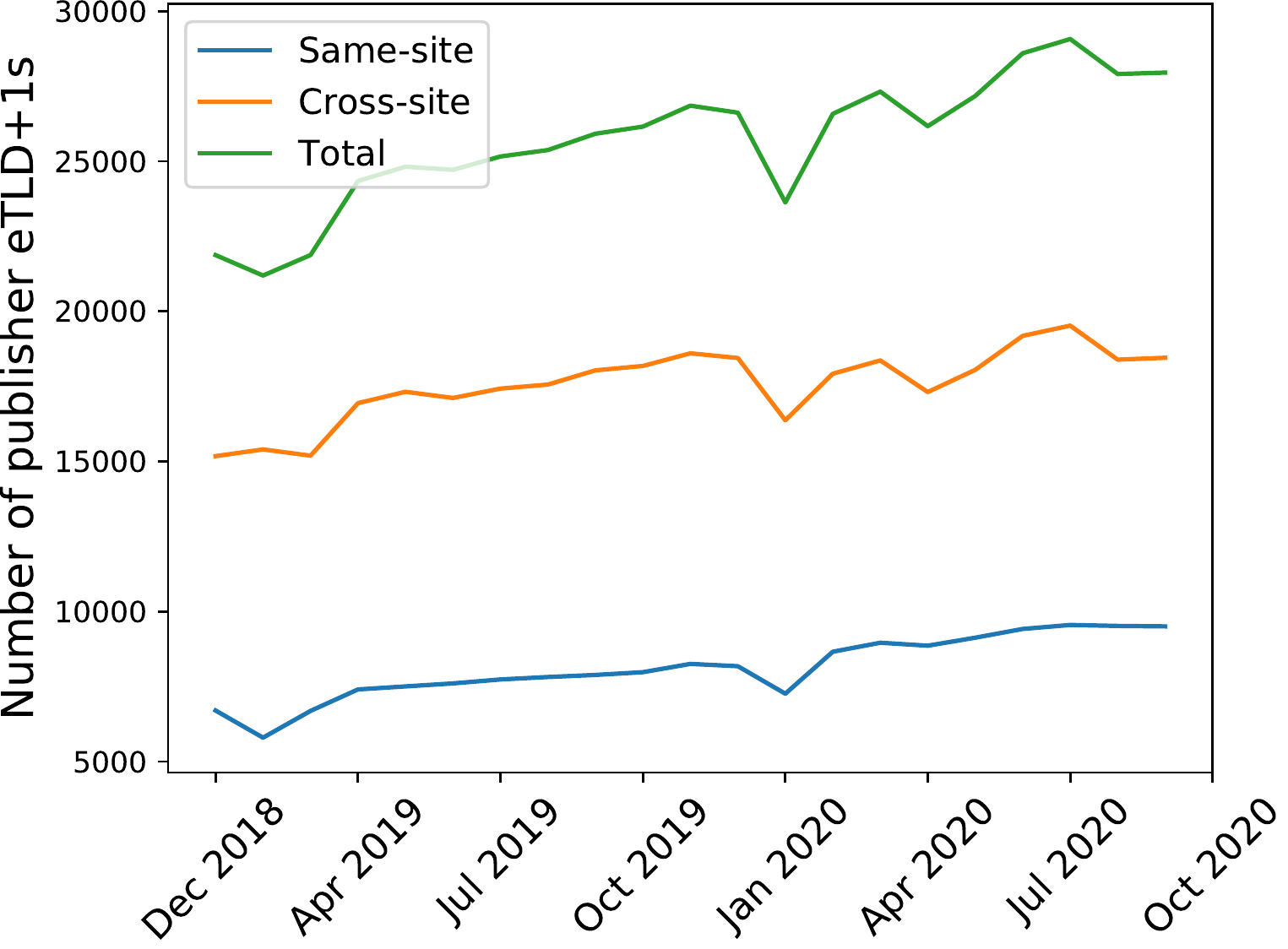} 
	\caption{Number of eTLD+1 domains that include CNAME-based tracking in a same-site and cross-site context.}
	\label{fig:number-of-customers-over-time}
\end{figure}

Figure~\ref{fig:number-of-customers-over-time} shows the total number of publisher eTLD+1s using CNAME-based tracking, either in a same-site or cross-site context.
The sudden drop in number of cross-site inclusions of CNAME trackers in October 2019 is mainly due to a single tracker (Adobe Experience Cloud).
We suspect it is related to changes it made with regard to CCPA regulations (the HTTP Archive crawlers are based in California)~\cite{omnitureccpa}.
In general, we find that the number of publisher sites that employ CNAME-based tracking is gradually increasing over time.

\begin{figure}
	\centering
	\includegraphics[width=0.9\linewidth]{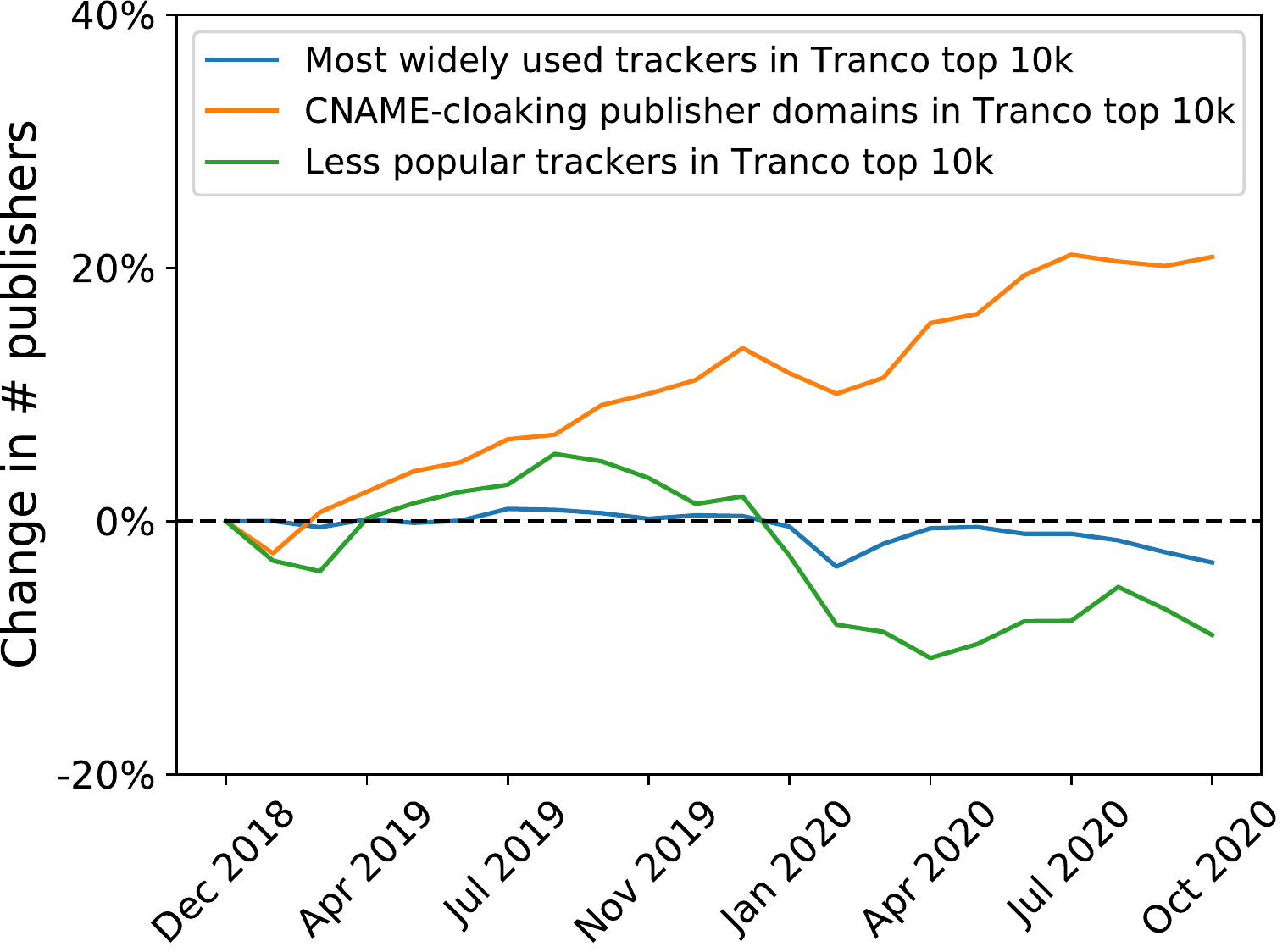} 
	\caption{Relative percentage, based on the state as of December 2018, of the number of publishers of popular and less popular trackers and CNAME-based trackers.}
	\label{fig:comparison-over-time}
\end{figure}

To further explore the evolution of the adoption of CNAME-based tracking, we compare it to the evolution of third-party tracking on the web. 
More specifically, for the ten most popular tracking companies according to WhoTracks.me~\cite{karaj2018whotracks}, and fifteen randomly selected less popular trackers with between 50 and 15,000 publishers as of October 2020 (similar to the customer base we observed for the CNAME-based trackers), we determined the number of publishers in the Tranco top 10k list\footnote{\url{https://tranco-list.eu/list/Z7GG/10000}}, between December 2018 and October 2020.
To this end we used the EasyPrivacy block list, and only used the rules that match the selected trackers.
For the three cases (popular trackers, less popular trackers and CNAME-based trackers) we computed the relative increase or decrease in number of publishers for the Tranco top 10k websites.
As the point of reference, we take the first entry of our dataset: December 2018.
The relative changes in the number of publishers are shown in Figure~\ref{fig:comparison-over-time}, and indicate that the customer base of less popular trackers declines whereas popular trackers retain a stable customer base.
This is in line with the findings of a study by Cliqz and Ghostery~\cite{whotracksmegdpr}.
Our results clearly show that compared to third-party trackers, the CNAME-based trackers are rapidly gaining in popularity, with a growth of 21\% over the past 22 months (compared to a change of $-3\%$ for popular trackers and $-8\%$ for less popular trackers).

\subsection{Method evaluation}
\label{sec:method_validation}
In this section, we evaluate the method we used to detect CNAME-based tracking throughout time for correctness and completeness.
For this analysis, we make use of historical DNS data provided by Rapid7  \cite{rapid7fdns}. 
We try to determine both the web pages that were incorrectly considered to be using CNAME-based tracking, as well as publishers that we might have missed by using our method.

\textbf{Correctness}
To assess the correctness of our approach, we looked for subdomains that we considered to be using CNAME tracking for each month of our analysis (December 2018 until October 2020), but that did not have a CNAME record pointing to a tracker in the corresponding month in the historical Rapid7 DNS dataset.
We found 81 publishers, 0.46\% of the 17,633 publishers that we determined over the whole period, that could potentially be labeled incorrectly.
Upon a closer examination, we find that all of these 81 publishers were in fact correctly marked as such.\\
These 81 publishers can be divided in three major groups based on the same reason that caused the mismatch in the datasets.
\textit{First}: Because of the timing difference between the HTTP Archive dataset and the Rapid7 dataset, the tracking domain of 21 publishers did not yet appear in the Rapid7 DNS dataset in the first month of starting to use CNAME-based tracking.
\textit{Second}: We found that 15 CNAME-based tracking domains incorrectly configured their DNS records, causing them to send tracking requests to an non-existent or typo domain.
For instance, several CNAME records pointed to a \texttt{.207.net} domain instead a \texttt{.2o7.net} domain.
\textit{Third}: We found 42 publisher tracking subdomains that did not have a CNAME record pointing to a known tracking domain. Instead, it pointed to another domain that would still resolve to the same IP address used by the tracker.
This occurs when the tracker adds a new tracking domain but the publisher that included it did not yet update their CNAME records.
For example, we observe nine publisher subdomains that have a CNAME record pointing to \texttt{.ca-eulerian.net}, whereas the currently used domain is \texttt{.eulerian.net}.
On the other hand, as of October 2020, Adobe Experience Cloud added a new tracking domain, namely \texttt{data.adobedc.net}; in the dataset of this month we found 33 tracking subdomains that already started referring to it.
As our method is agnostic of the domain name used in the CNAME record of the publisher subdomain (the domain name may change over time), it can detect these instances, in contrast to an approach that is purely based on CNAME records.
Finally, for the remaining three publishers, we found that a DNS misconfiguration on the side of the publisher caused the CNAME record to not correctly appear in the Rapid7 dataset.
Although tracking requests were sent to the tracking subdomain, these subdomains would not always resolve to the correct IP address, or return different results based on the geographic location of the resolver.\\
As a result, we conclude that all of the publishers were correctly categorized as using CNAME-based tracking.
Moreover, our method is robust against changes in tracking domains used by CNAME trackers.

\textbf{Completeness}
We evaluate the completeness of our method by examining domain names that we did not detect as publishers, but that do have a CNAME record to a tracking domain.
Our detection method uses an accumulating approach starting from the most recent month's data (October 2020) and detecting CNAME-based tracking for each previous month, based on the current month's data.
For this reason, we only consider publisher subdomains that we might have missed in the final month of our analysis (December 2018), where the missed domains error would be most notable. 
Out of the 20,381 domain names that have a CNAME record in the Rapid7 dataset pointing to a tracking domain, 12,060 (59.2\%) were not present in the HTTP Archive dataset.
From the remaining domain names, 7,866 (38.6\%) were labeled as publishers by us, leaving 455 (2.2\%) domain names that we \emph{potentially} missed as a consequence of using our method.
After examining the HTTP Archive dataset for these domains, we find that for 195 hostnames the IP address is missing in the dataset.
For the remaining 260 domains, we find that the majority (196) does not send any tracking-related request to the tracker, which could indicate that the tracking service is not actively being used.
For 41 domain names, we find that the sent requests do not match our request pattern, and further examination shows that these are in fact using another service, unrelated to tracking, from one of the providers.
The remaining 22 domain names were missed as publishers in our method since these resolved to an IP address that was not previously used for CNAME-based tracking. \\
Our results show that relying solely on DNS data to detect CNAME-based tracking leads to an overestimation of the number of publishers.
Furthermore, our method missed only 0.28\% of CNAME-based tracking publishers due to irregularities in the set of IP addresses used by CNAME-based tracking providers.
A downside of our method is that it cannot automatically account for changes of the request signature used by CNAME trackers throughout time.
However, we note that in the analysis spanning 22 months, we did not encounter changes in the request signature for any of the 13 trackers.

\textbf{Tracker domain ownership}
Lastly, we verify whether the ownership of the IP-addresses used by the thirteen trackers changes throughout time. To achieve this, we examine PTR records of the IP-addresses used for tracking in December 2018 and check whether the owner company of the resulting domains has changed since then, by using Rapid7's reverse DNS dataset \cite{rapid7rdns} and historical WHOIS data \cite{whoxy}. 
We find that all of the IP addresses point to domains owned by the corresponding tracker. Furthermore, for 7 trackers, the ownership of the tracking domains has not changed since December 2018. 6 trackers had redacted their WHOIS information due to privacy,  out of which 1 was not updated throughout our measurement period. The other 5 have been updated recently and therefore we cannot conclude that their owner has remained the same. We do suspect this is the case however, since all of the domains were owned by the corresponding tracker before the details became redacted.

\subsection{Effects on third-party tracking}
In order to gather more insight on the reasons as to why websites adopt CNAME-based tracking, we performed an additional experiment.
We posed the hypothesis that if the number of third-party trackers employed by websites decreases after they started using the CNAME-based tracking services, this would indicate that the CNAME-based tracking is used as a replacement for third-party tracking.
A possible reason for this could be privacy concerns: without any anti-tracking measures, third-party tracking allows the tracker to build profiles of users by following them on different sites, whereas CNAME-based tracking only tracks users on a specific site (assuming that the tracker acts in good faith).
Conversely, if the number of third-party trackers remains stable or even increases, this would indicate that CNAME-based tracking is used in conjunction with third-party tracking, e.g.\ to still obtain information on users that employ anti-tracking measures.

\begin{figure}
	\centering
	\includegraphics[width=\linewidth]{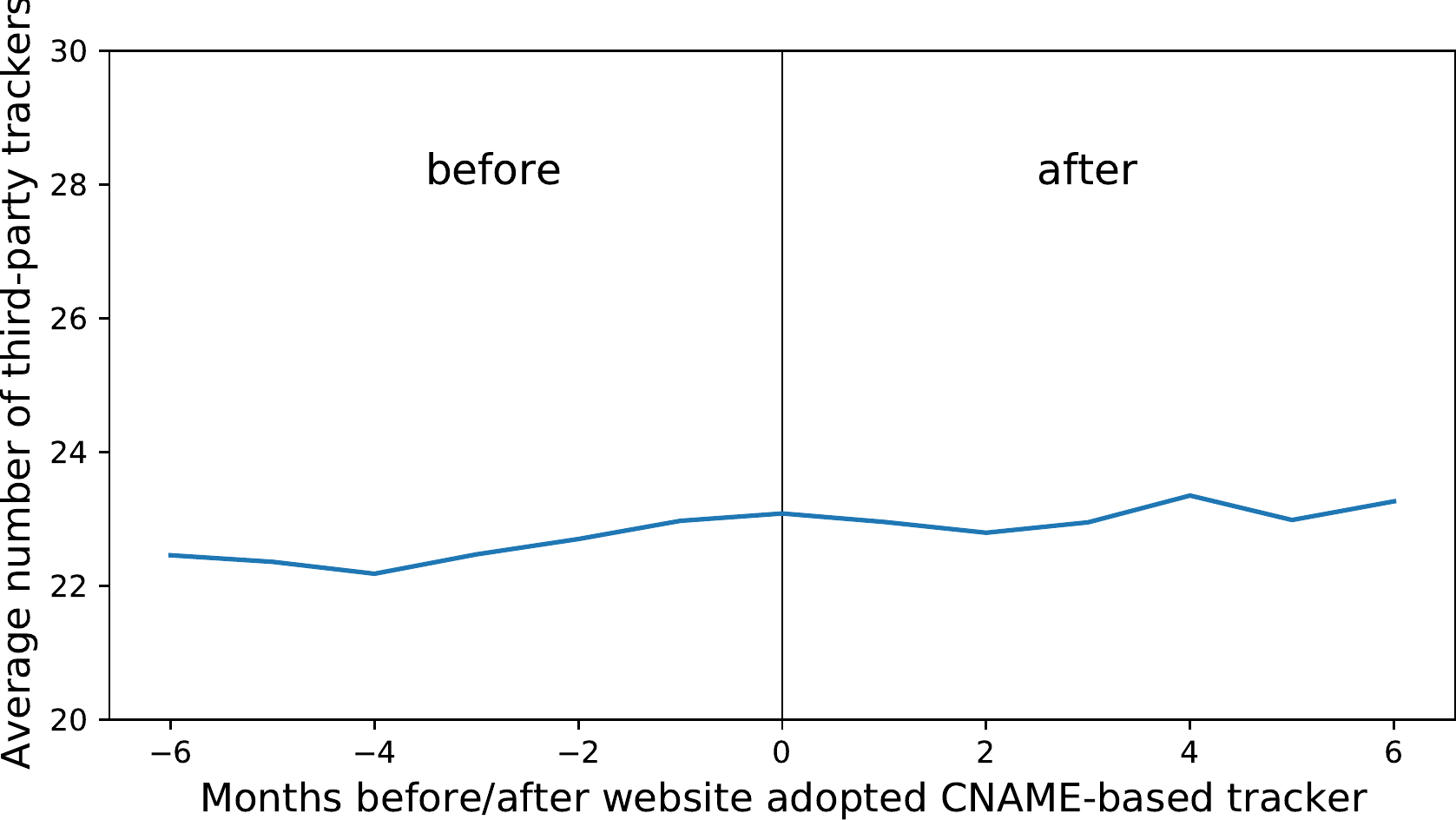} 
	\caption{Number of third-party trackers adopted by publishers in the six months before and after they adopted a CNAME-based tracker.}
	\label{fig:third-parties-over-time}
\end{figure}

To measure the evolution of the number of third-party trackers on publisher sites that recently adopted CNAME-based tracking, we again use the measurements ranging between December 2018 and October 2020 from the HTTP Archive dataset.
We consider a publisher website including a CNAME tracker to be a \emph{new} if for six consecutive months it did not refer to this tracker through a CNAME record on a subdomain, and then for the following six months always included a resource from this tracker.
In total we found 1,129 publishers at in the duration of our analysis started using CNAME tracking.
For these publishers, we determined the number of third-party trackers based on the EasyPrivacy blocklist for the six months before and after the time the publishers adopted CNAME-based tracking.
The average number of third-party trackers over this time period is shown in Figure~\ref{fig:third-parties-over-time}.
We find that the adoption of CNAME-based tracking services does not significantly affect the third-party trackers that are in use, indicating that these CNAME-based trackers are used to complement the information obtained from other trackers.

\section{Implications of first-party inclusion}
\label{sec:implications-of-cname-tracking}
In this section we investigate how CNAME-based tracking can expand a website’s attack surface.
Since CNAME-based trackers are included in a same-site context, there may be additional security risks compared to third-party trackers. For instance, privacy-sensitive information, e.g.\ contained in cookies, may be inadvertently sent to the tracker, posing increased threats for users.

\subsection{Transport security}
When visiting a website that employs CNAME-based tracking, various types of requests are made to the tracker-controlled subdomain.
We find that most commonly, the web page makes a request to report analytics data, typically via an asynchronous request or by creating an (invisible) <img> element.
Additionally, we find that in most cases the tracking script is also included from the CNAME subdomain. 
To ensure that a man-in-the-middle attacker cannot read or modify the requests and responses, a secure HTTPS connection is required.
Based on the HTTP Archive dataset from July 2020, we find that the vast majority (92.18\%) of websites that use CNAME-based tracking support TLS, and in almost all cases the tracker requests are sent over secure connections.
Nevertheless, we did identify 19 websites where active content, i.e.\ HTML or JavaScript, was requested from the tracker over an insecure connection.
Although most modern browsers block these requests due mixed content policies, users with outdated browsers would still be susceptible to man-in-the-middle attacks.

On 72 websites we found that an analytics request sent to a CNAME-based tracker was sent over HTTP while the web page was loaded over HTTPS.
In this case, the request is not blocked but instead the browser warns the user that the connection is insecure.
Because this is a same-site request (as opposed to a cross-site request as would be the case with third-party tracking), cookies that are scoped to the eTLD+1 domain, and that do not contain the \texttt{Secure} attribute, are attached to this request.
Consequently these potentially identifying cookies can be intercepted on by network eavesdroppers.
Furthermore an attacker could exploit unencrypted HTTP responses.
Specifically, the adversary could inject arbitrary cookies in \texttt{Set-Cookie} headers to
%responses
%to set a cookie with an arbitrary value on the including website, effectively 
launch a session-fixation attack~\cite{kolvsek2002session,schrank2010session}.
In the remainder of this section, we explore the privacy and security threats associated with including the tracker as first party in more detail.

\subsection{Tracker vulnerabilities: case studies}
To further explore how the security of websites and their visitors is affected by including a CNAME-based tracker, we performed a limited security evaluation of the trackers that are included on publisher websites.
For up to maximum 30 minutes per tracker, we analyzed the requests and responses to/from the CNAME subdomain for client-side web vulnerabilities.
In most cases, we found that only a single request was made, and an empty response was returned.
Despite the time-limited nature of our analysis, we did identify vulnerabilities in two different trackers that affect all publishers that include them.
We reported the vulnerabilities to the affected trackers and actively worked with them to mitigate the issues.
Unfortunately, in one instance the tracker did not respond to repeated attempts to report the vulnerability, leaving hundreds of websites exposed.
We still hope to be able to contact this vendor through one of their customers.

\subsubsection{Vulnerability 1: session fixation} The first vulnerability is caused by the tracker's functionality to extend the lifetime of first-party advertising and analytics cookies, such as Facebook's \texttt{\_fbp} cookie or the \texttt{\_ga} cookie by Google Analytics. 
Because these cookies are set by a cross-site script through the \texttt{document.cookie} API, Safari's ITP 
%and Firefox' ETP 
limits their lifespan to seven days~\cite{safari-itp-21}.
To overcome these limits, the tracker provides a specific endpoint on the CNAME subdomain that accepts a POST request with a JSON payload containing the cookie names and values whose lifetime should be extended.
In the response, the tracker's server includes several \texttt{Set-Cookie} headers containing the tracking cookies.
Consequently, these cookies are no longer set via the DOM API and would have an extended lifetime under Safari's ITP policies for cookies. We note that this circumvention is disabled as of late 2020, thanks to Safari's recent ITP update targeting CNAME-based trackers. This update caps the lifetime of HTTP cookies from CNAME trackers to seven days, which matches the lifetime of cookies set via JavaScript~\cite{safari-cname-defense}.

We found that the tracker endpoint did not adequately validate the origin of the requests, nor the cookie names and values.
Consequently, through the functionality provided by the tracker, which is enabled by default on all the websites that include the tracker in a first-party context, it becomes possible to launch a session-fixation attack.
For example, on a shopping site the attacker could create their own profile and capture the cookies associated with their session.
Subsequently, the attacker could abuse the session-fixation vulnerability to force the victim to set the same session cookie as the one from the attacker, resulting in the victim being logged in as the attacker.
If at some point the victim would try to make a purchase and enter their credit card information, this would be done in the attacker's profile.
Finally, the attacker can make purchases using the victim's credit card, or possibly even extract the credit card information.

The impact of this vulnerability highlights the increased threat surface caused by using the CNAME-based tracking scheme.
If a third-party tracker that was included in a cross-site context would have the same vulnerability, the consequences would be negligible.
The extent of the vulnerability would be limited to the setting of an arbitrary cookie on a tracking domain (as opposed to the first-party visited website) which would have no effect on the user.
However, because in the CNAME-tracking scheme the tracking domain is a subdomain of the website, cookies set with a \texttt{Domain} attribute of the eTLD+1 domain (this was the default in the detected vulnerability), will be attached to all requests of this website and all its subdomains.
As a result, the vulnerability does not only affect the tracker, but introduces a vulnerability to all the websites that include it.

\subsubsection{Vulnerability 2: cross-site scripting} The second vulnerability that we identified affects publishers that include a different tracker, and likewise it is directly related to tracker-specific functionality.
In this case, the tracker offers a method to associate a user's email address with their fingerprint (based on IP address and browser properties such as the User-Agent string).
This email address is later reflected in a dynamically generated script that is executed on every page load, allowing the website to retrieve it again, even if the user would clear their cookies.
However, because the value of the email address is not properly sanitized, it is possible to include an arbitrary JavaScript payload that will be executed on every page that includes the tracking script.
Interestingly, because the email address is associated with the user's browser and IP fingerprint, we found that the payload will also be executed in a private browsing mode or on different browser profiles.
We tested this vulnerability on several publisher websites, and found that all could be exploited in the same way.
As such, the issue introduced by the tracking provider caused a persistent XSS vulnerability in several hundreds of websites.

\subsection{Sensitive information leaked to CNAME-based trackers}
\label{sec:privacy-implications}
CNAME-based trackers operate on a subdomain of publisher websites.
It is therefore possible that cookies sent to the tracker may contain sensitive information, such as personal information (name, email, location) and authentication cookies, assuming these sensitive cookies are scoped to the eTLD+1 domain of the visited website (i.e. \texttt{Domain=.example.org}).
Furthermore, it is possible that websites explicitly share personal information with the CNAME-based trackers in order to build a better profile on their users.

To analyze the type of information that is sent to trackers and to assess the frequency of occurrence, we performed a manual experiment on a random subset of publishers.
Based on data from a preliminary crawl of 20 pages per website, we selected up to ten publisher websites per tracker that had at least one HTML form element with a password field.
We limited the number of websites in function of the manual effort required to manually register, login, interact with it, and thoroughly analyze the requests that were sent.
We looked for authentication cookies (determined by verifying that these were essential to remain logged on to the website), and personal information such as the name and email that was provided during the registration process.

Out of the 103 considered websites, we were able to successfully register and log in on 50 of them.
In total, we found that on 13 of these websites sensitive information leaked to the CNAME tracker.
The leaked information included the user's full name (on 1 website), location (on 2 websites), email address (on 4 websites, either in plain-text or hashed), and the authentication cookie (on 10 websites). 
We note that such leaks are the result of including the trackers in a first-party context.
Our limited study indicates that the CNAME tracking scheme negatively impacts users' security (authentication cookie leaks) and privacy (personal data leaks).

\subsection{Cookie leaks to CNAME-based trackers}
\label{sec:cookie-leak-automated-experiment}

Next we perform an automated analysis to investigate cookies that are inadvertently sent to CNAME trackers.
We conducted an automated crawl on June 7, 2020 of 8,807 websites that we, at that time, identified as using CNAME-based tracking following the methodology outlined in Section~\ref{sec:tracking-customers}.
In this crawl, we searched for cookies sent to the CNAME subdomain while excluding the cookies set by the CNAME tracker itself (either through its subdomain or its third-party domains).

\textbf{The crawler}
We built our crawler by modifying the DDG Tracker Radar Collector \cite{ddg}, a Puppeteer-based crawler that uses the Chrome DevTools Protocol (CDP).
We extended the crawler by adding capabilities to capture HTTP request cookies, POST data, and document.cookie assignments.
DDG Tracker Radar Collector uses the Chrome DevTools Protocol to set breakpoints and capture the access to the Web API methods and properties that may be relevant to browser fingerprinting and tracking (e.g.\ document.cookie).
We used this JavaScript instrumentation to identify scripts that set cookies using JavaScript.

For each website, we loaded the homepage using a fresh profile.
We instructed the crawler to wait ten second on each website, and then reload the page.
This allowed us to capture the leaks of cookies that were set after the request to the CNAME-based tracker domain.
We also collected HTTP headers, POST bodies, JavaScript calls, and cookies from the resulting profile.
When crawling, we used a Safari User-Agent string, as we found at least one CNAME-based tracker (Criteo) employing first-party tracking for Safari users only.

\textbf{Data analysis}
To identify the cookie leaks, we first built the list of cookies sent to the CNAME subdomain.
From the resulting list, we excluded session cookies, short cookies (less than 10 characters), and cookies that contain values that occur on multiple visits (to exclude non-uniquely identifying cookies).
To determine the latter, we first built a mapping between the distinct cookie values and the number of sites they occur on.

Next, we identified the \emph{setter} of the cookies.
First, we searched the cookie name and value in \texttt{Set-Cookie} headers in HTTP responses.
When the cookie in question was sent in the corresponding request, we excluded its response from the analysis.
For JavaScript cookies, we searched for the name-value pair in assignments to \texttt{document.cookie} using the JavaScript instrumentation data.
We then used the JavaScript stack trace to determine the origin of the script.
After determining the setter, we excluded cookies set by the CNAME-based tracker itself.

\textbf{Leaks in HTTP Cookie headers}

We identified one or more cookie leaks on 7,377 sites (95\%) out of the 7,797 sites where we could identify the presence of at least one CNAME-based tracker.
Table~\ref{tab:cookie-leak} shows the five origins with most cookies leaked to CNAME-based trackers.
The overwhelming majority of cookie leaks (31K/35K) are due to third-party analytics scripts setting cookies on the first-party domain.

The leakage of first-party cookies containing unique IDs may not reveal any additional information to CNAME-based trackers, since these trackers may already have an ID for the users in their own cookies.
However, cookies containing other information such as ad campaign information, emails, authentication cookies may also leak to the CNAME-based trackers (as shown in Section~\ref{sec:privacy-implications}).
Moreover, our analysis found that on 4,006 sites, a cookie set by a third-party domain is sent to the CNAME-based tracker's subdomain.
3,898 of these sites are due to Pardot, which sets the same cookie on its first-party subdomain and its third-party domain.
To set the same cookie on both domains, Pardot sends its unique ID in a URL parameter called~\texttt{visitor\_id} to its first-party subdomain. 

\begin{table}[]
\centering
\caption{Five origins with most leaked cookies to CNAME-based trackers. The right column indicates the number of distinct sites cookies we observed one or more cookie leaks set by the scripts from these origins.}
\begin{adjustbox}{max width=\textwidth}

\begin{tabular}{llr}
\toprule
\textbf{Cookie origin} & \textbf{Purpose} & \textbf{\shortstack{Num.\ of \\ distinct sites}} \\ \midrule
www.google-analytics.com & Analytics      & 5,970 \\
connect.facebook.net     & FB Pixel       & 3,287 \\
www.googletagmanager.com & Tag management & 2,376 \\
bat.bing.com             & Advertising    & 1,182 \\
assets.adobedtm.com      & Tag management & 887  \\ \bottomrule
\end{tabular}
\end{adjustbox}
\label{tab:cookie-leak}
\end{table}

\textbf{Leaks in POST request bodies}
While we accept and do not rule out that cookie leaks may often happen inadvertently, i.e.\ without the knowledge or the cooperation of the CNAME trackers, when browsers send cookies with a matching domain to the tracker, this picture is not always so straight-forward.
Namely, we identified and investigated two other types of cookie leaks that involve more active participation by the CNAME trackers.
First, we studied cookie values sent in the POST request bodies, again excluding the cookies set by the CNAME tracker itself, and session cookies and cookies that occur on multiple sites, as described above.
We found that 166 cookies (on 94 distinct sites) set by another party were sent to a CNAME tracker's subdomain in a POST request body.
The majority of these cases were due to TraceDock (46 sites) and Adobe Experience Cloud (30 sites), while 
Otto Group and Webtrekk caused these cookie leaks on 11 and seven sites respectively.

We used the request ``initiators'' field to identify the \emph{senders} of the requests.
The ``initiators'' field contains the set of script addresses that triggered an HTTP request, derived from JavaScript stack traces.
In 78 of the 166 instances, the CNAME subdomain or the tracker's third-party domains were among the initiators of the POST request. 
In the remaining cases, the CNAME tracker's script was served on a different domain (e.g. Adobe Experience Cloud, assets.adobedtm.com), a different subdomain that also belongs to the CNAME tracker (e.g.\ Otto Group uses tp.xyz.com subdomain for its scripts and te.xyz.com for the endpoint), or the request was triggered by a tag manager script,
or a combined script that contains the CNAME tracker's script.

The cookies sent in the POST bodies indicate that certain CNAME tracker scripts actively read and exfiltrate cookies they may access on first party sites.
Although the content of the cookies may not always reveal additional information, our manual analysis presented above revealed sensitive information such as email addresses, authentication cookies and other personal information is leaking to the CNAME trackers.

\textbf{Leaks in request URLs}
Next we investigate the cookies sent to CNAME tracker subdomains in the request URLs.
To detect such leaks we searched for cookies in the request URLs (and URL-decoded URLs) excluding the scheme and the hostname.
We excluded the same set of cookies as the previous two analyses -- cookies set by CNAME tracker itself, short cookies,  session cookies and cookies with non-unique values.

We found 1,899 cookie leaks in request URLs to CNAME subdomains on 1,295 distinct sites. 
1,566 of the cookies were sent to Adobe Experience Cloud's subdomain, while Pardot's and Eularian's subdomains received 130 and 101 cookies, respectively.
In addition, in 4,121 cases (4,084 sites), a cookie set by Pardot's third-party domain was sent to its CNAME subdomain, confirming the finding above that Pardot syncs cookies between its third-party domain and its CNAME subdomain.
Overall, in 378 cases the leaked cookie was set by a third-party domain, indicating that cookies were synced or simply exchanged between the domains.

Our automated analysis of cookie leaks, in combination with the deeper manual analysis presented above indicates that passive and active collection of cookies by the CNAME trackers is highly prevalent and have severe privacy and security implications including the collection of email addresses, unique identifiers and authentication cookies.
Further, our results show that certain CNAME-based trackers use third-party cookies for cross-site tracking and at times receive cookies set by other third-party domains, allowing them to track users across websites.

\section{Discussion}
\label{sec:countermeasures-and-circumvention}
CNAME-based tracking exists for several years now. Our analysis shows that recently it is rapidly gaining in popularity, especially on frequently-visited websites.
In this section we explore the current countermeasures against this form of tracking, and discuss their effectiveness and potential circumvention techniques that trackers may use in the future.

\textbf{Countermeasures}
In response to a report that a tracker was using CNAMEs to circumvent privacy blocklists\footnote{\url{https://github.com/uBlockOrigin/uBlock-issues/issues/780}}, uBlock Origin released an update for its Firefox version that thwarts CNAME cloaking~\cite{ublockrelease}.
The extension blocks requests to CNAME trackers by resolving the domain names using the \texttt{browser.dns.resolve} API to obtain the last CNAME record (if any) before each request is sent.
Subsequently, the extension checks whether the domain name matches any of the rules in its blocklists, and blocks requests with matching domains while adding the outcome to a local cache.
Although uBlock Origin has a version for Chromium-based browsers, the same defense cannot be applied because Chromium-based browser extensions do not have access to an API to perform DNS queries.
%As such, at the time of this writing, it is technically impossible for these extensions to block requests to trackers that leverage CNAME records to avoid detection.
As we explain in Section~\ref{sec:cname-tracking}, uBlock Origin for Chrome, which does not have a defense for CNAME-based tracking, still manages to block several trackers.
This is because the requests to the trackers matched an entry of the blocklist with a URL pattern that did not consider the hostname.
Unfortunately, it is fairly straightforward for the tracker to circumvent such a fixed rule-based measure, e.g.\ by randomizing the path of the tracking script and analytics endpoint, as is evidenced by the various trackers that could only be blocked by the uBlock Origin on Firefox.
An alternative strategy for browser extensions that do not have access to a DNS API could be to analyze the behavior or artifacts of tracking scripts.
However, the tracker's code could be dynamic and include many variations, making detection arduous and performance-intensive.
%As such, blocking CNAME-based trackers is highly challenging for browser extensions that do not have access to sufficient information to reveal the actual party that a request is made to.

Thanks to the increasing attention to CNAME-based tracking,
Safari and Brave recently followed uBlock Origin's suit, and implemented countermeasures against CNAME-based tracking.
Safari limited the expiry of cookies from CNAME trackers to seven days, which is the same limit they use for all cookies set by scripts~\cite{safari-cname-defense}.
Brave, on the other hand, started recursively checking for CNAME records of the network requests against their blocklists~\cite{brave-cname-defense}.
Mozilla is working on implementing a similar defense in Firefox~\cite{firefox-cname-defense}.

Other tracking countermeasures include DNS sinkholes that return a false IP address, (e.g.\ 127.0.0.1) when the domain name matches an entry from the blocklist.
As this type of countermeasure work at the DNS level, it considers all the intermediary resolutions to CNAME records, and effectively blocks the domains that match a blocklist.
Examples of DNS-based tools that adopted defenses against CNAME cloaking include NextDNS~\cite{nextdns2019cnamearticle}, AdGuard~\cite{adguardcnamedefense}, and Pi-hole~\cite{piholecnamedefense}.

\textbf{Circumvention}
Both anti-tracking solutions, i.e.\ browser extensions and DNS resolvers, rely on blocklists, and can thus only block trackers whose domain names are on the list.
Updating CNAME records using
randomized domain names may bypass these blocklists.
%This provides a first avenue for circumvention for the trackers: by randomizing the domain names that is referred to in CNAME records, it would become infeasible to rely on a domain-based blocklist.
However, this requires publishers to frequently update their CNAME records, which may be impractical for many websites.
%would mean that every time the tracker changes domains, all the publishers that include it would need to update their CNAME record, making it largely impractical.
Another circumvention option is to directly refer to the IP address of the tracker through an A record instead of a CNAME record.
We found the pool of IP addresses used by CNAME-based trackers to be relatively stable over time, and in fact found that several (35) publishers already use this method.
At the time of this writing, using IP addresses (and A records) circumvents blocklists, which do not use IP addresses to identify trackers.
%but this can be easily defended against by adding the IP addresses to the blocklist.

While IP addresses can be added to blocklists, changing IP addresses as soon as they are added to blocklists would be practically infeasible, as it requires all publishers to update their DNS records.
Nevertheless, a tracker could request their publishers to delegate authority for a specific subdomain/zone to the tracker by setting an NS record that points to the tracker.
As such, the tracker could dynamically generate A record responses for any domain name within the delegated zone, and thus periodically change them to avoid being added to blocklists.
For anti-tracking mechanisms to detect this circumvention technique, this would require obtaining the NS records to determine whether they point to a tracker.
Although it may be feasible to obtain these records, it may introduce a significant overhead for the browser extensions and DNS-based anti-tracking mechanisms.

In general, as long as the anti-tracking mechanism can detect the indirection to the third-party tracker, it is possible to detect and block requests to the tracker, albeit at a certain performance cost.
Trackers could try to further camouflage their involvement in serving the tracking scripts and collecting the analytics information.
For instance, they could request the publishers that include tracking scripts to create a reverse proxy for a specific path that points to the tracker, which could be as easy as adding a few lines in the web server configuration, or adjusting the settings of the CDN provider.
In such a situation, the tracking-related requests would appear, from a user's perspective, to be sent to the visited website, both in terms of domain name as well as IP address.
Thus, current tracking defenses would not be able to detect or block such requests.
As the perpetual battle between anti-tracking mechanisms and trackers continues, as evidenced by the increasing popularity of CNAME-based tracking, we believe that further empirical research on novel circumvention techniques is warranted.

\textbf{Limitations}
As stated in Section~\ref{sec:historical-evolution}, the method we use to detect CNAME-based tracking in historical data cannot account for changes in the request signature used by trackers. In practise, these signatures remained the same during our measurement period. Furthermore, part of the experiments we conducted in Section~\ref{sec:implications-of-cname-tracking} required substantial manual analysis, making it infeasible to perform on a larger set of websites. 

\section{Related work}

In 2009, Krishnamurthy and Wills provided one of the first longitudinal analyses of user information flows to third-party sites (called \textit{aggregators})~\cite{krishnamurthy2009privacy}.
The authors also observed a trend of serving third-party tracking content from first-party contexts, pointing out the challenges for countermeasures based on blocklists.
Meyer and Mitchell studied the technology and policy aspects of third-party tracking~\cite{mayer2012third}. Englehardt and Narayanan \cite{englehardt2016online} 
measured tracking on Alexa top million websites using OpenWPM and discovered new fingerprinting techniques such as AudioContext API-based fingerprinting.

The CNAME tracking scheme was mentioned anecdotally by Bau in 2013 \cite{bau2013promising}, but the authors did not focus on the technique specifically.
To our knowledge, the first systematic analysis of the CNAME scheme used to embed third-party trackers in first-party content is the work of Olejnik and Casteluccia \cite{olejnik2014analysis}, 
in which they identified this special arrangement as part of the real-time bidding setup.
The authors also reported leaks of first-party cookies to such third parties. In our paper, we extensively expand such analyses.
Although cookies were most commonly used for cross-site tracking, more advanced mechanisms have been deployed by websites and studied by the researchers. Browser fingerprinting \cite{eckersley2010unique}, where traits of the host \cite{yen2012host}, system, browser and graphics stack~\cite{mowery2012pixel} are extracted to identify the user is one of the stateless tracking vectors that does not need cookies to operate.
Fingerprinting on the web was measured at scale by Acar et al. \cite{acar2013fpdetective, Acar:2014:WNF:2660267.2660347}, Nikiforakis et al.\cite{nikiforakis2013cookieless},
and Englehardt and Narayanan~\cite{englehardt2016online}. Combining multiple tracking vectors at the same time may give rise to supercookies or evercookies, as demonstrated first by Samy Kamkar \cite{kamkar2010evercookie}.
Over the years, many information exfiltration or tracking vectors have been studied, including Cache Etag HTTP header~\cite{ayenson2011flash}, WebSockets \cite{bashir2018tracking}, ultrasound beacons \cite{mavroudis2017privacy}, and fingerprinting sensors calibrations on mobile devices with sensors \cite{zhang2019sensorid}.

Similar to these studies we measure the prevalence of a tracking mechanism that tries to circumvent existing countermeasures. However our work uses novel methods to identify CNAME-based trackers in historical crawl data, allowing us to perform a longitudinal measurement.

In concurrent work, Dao et al.\ also explored the ecosystem of CNAME-based trackers~\cite{dao2020cnamecloaking}.
Based on a crawl of the Alexa top 300k, they find 1,762 CNAME-based tracking domains as of January 2020, which are detected by matching the CNAME domain with EasyPrivacy.
In our work, we detected 9,273 sites that leverage CNAME-based tracking in a same-site context and an additional 19,226 websites that use it in a cross-site context.
We rely on an approach that combines historical DNS records (A records) with manually constructed fingerprints.
The latter is used to filter out any potential false positives that may be caused by changes in the IP space ownership, or because the CNAME- or A-records may be used to other services of the same provider unrelated to tracking.
Based on the evaluation of our method in Section~\ref{sec:method_validation}, we find that it is important to use request-specific information to prevent incorrectly marking domains as using CNAME-based tracking.
Furthermore, relying on filter lists, and in particular on the eTLD+1 domains that are listed, could result in the inclusion of non-tracking domains, e.g.\ sp-prod.net is the second most popular tracker considered by Dao et al., but was excluded in our work as it is part of a ``Consent Management Platform'' that captures cookie consent for compliance with GDPR \cite{CMP}.
Additionally, filter lists may be incomplete, resulting in trackers being missed: for example, Pardot, the tracker we find to be most widely used, was not detected in prior work.
Consequently, relying on filter lists also prevents the detection of new trackers, this limitation is not applicable to our method.

Dao et al.\ also perform an analysis of the historical evolution of CNAME-based tracking, based on four datasets of the Alexa top 100k websites collected between January 2016 and January 2020.
As the used OpenWPM datasets do not include DNS records, the researchers rely on a historical forward DNS dataset provided by Rapid7~\cite{rapid7fdns}, which does not cover all domains over time.
By using the HTTP Archive dataset, which includes the IP address that was used, we were able to perform a more granular analysis, showing a more accurate growth pattern.
We also show that this growth is rapidly increasing, significantly outperforming third-party trackers with a comparable customer base.
Finally, to the best of our knowledge, we are the first to perform an analysis of the privacy and security implications associated with the CNAME-based tracking scheme.

\section{Conclusion}

Our research sheds light on the emerging ecosystem of CNAME-based tracking, a tracking scheme that takes advantage of a DNS-based cloaking technique to evade tracking countermeasures.
Using HTTP Archive data and a novel method, we performed a longitudinal analysis of the CNAME-based tracking ecosystem using crawl data of 5.6M web pages.
Our findings show that unlike other trackers with similar scale, CNAME-based trackers are becoming increasingly popular, and are mostly used to supplement ``typical'' third-party tracking services.
We evaluated the privacy and security threats that are caused by including CNAME trackers in a same-site context.
Through manual analysis we found that sensitive information such as email addresses and authentication cookies leak to CNAME trackers on sites where users can create accounts.
Furthermore, we performed an automated analysis of cookie leaks to CNAME trackers and found that cookies set by other parties leak to CNAME trackers on 95\% of the websites that we studied.
Finally we identified two major web security vulnerabilities that CNAME trackers caused.
We disclosed the vulnerabilities to the respective parties and have worked with them to mitigate the issues.
We hope that our research helps with addressing the security and privacy issues that we highlighted, and inform development of countermeasures and policy making with regard to online privacy and tracking. 

\newpage
\bibliographystyle{ACM-Reference-Format}
\bibliography{cname-references}
\appendix
\section{Acknowledgement}
\label{sec:acknowledgements}

This research is partially funded by the Research Fund KU Leuven, and by the Flemish Research Programme Cybersecurity with reference number VR20192203. We would like to thank Steve Englehardt and the reviewers for their constructive comments. Gunes Acar holds a Postdoctoral fellowship of the Research Foundation Flanders (FWO).
\section{Assisted detection}
\label{sec:assisted-detection}

First-party subdomains referring to third-parties are by no means exclusive to CNAME-based tracking: 
services such as CDNs rely on a similar setup. Many websites hosting various services utilize CNAMEs to connect website domains to third-party hosts. Furthermore, a variety of different kinds of services provide third-party content in a first-party context by using CNAME records. Examples include  Consent Management Providers or domain parking services and traffic management platforms.\\

In our approach to distinguish the various kinds of first-party services we collected features that 
help us characterize a resource. For each of the 120 services we considered, we measured the number of websites the first-party is active on, the number of different hostnames a request to the service originates from, and the number of unique paths occurring in requests to the service. Furthermore, we captured the body size of the response, 
its content type (i.e. an image, script, video or html resource) and the average number of requests per website using the service. 
Lastly, we detected the percentage of requests and websites that sent and received cookies from the service.\\

To measure the uniformity of the response sizes of potential first-party trackers we sorted the sizes in buckets, each bucket with a size of 100 bytes. We then considered the number of buckets as a possible feature for distinction between different kinds of services. A low number of buckets would indicate that the service has a similar response to each request (e.g. the same script) which would increase the likelihood of the service being a tracker. 

After manually visiting the websites of each of the considered services, we were able to classify them in three different categories: \textit{trackers}, \textit{Content Distribution Networks} (CDNs) and \textit{other}. Any service that did not mention being explicitly a CDN or a tracker on their website, was categorized as ``other''.

\begin{figure}
	\begin{subfigure}[b]{0.45\linewidth}
		\includegraphics[width=\linewidth]{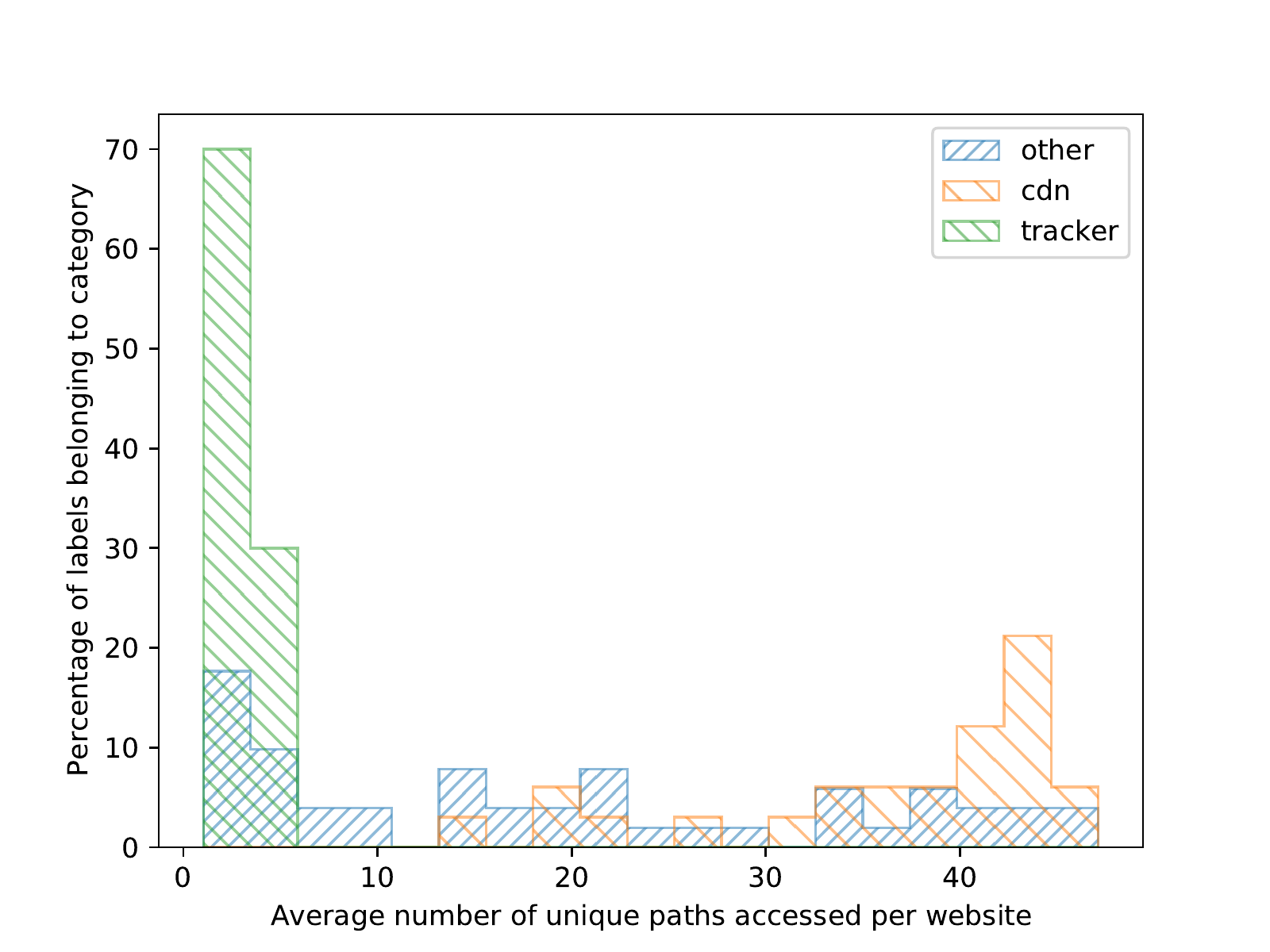}
		\caption[Network2]
		{{\small Distribution of the average number of unique paths per website}}    
		\label{fig:avg_request_paths_per_page}
	\end{subfigure}
	\hfill
	\begin{subfigure}[b]{0.45\linewidth}  
		\includegraphics[width=\linewidth]{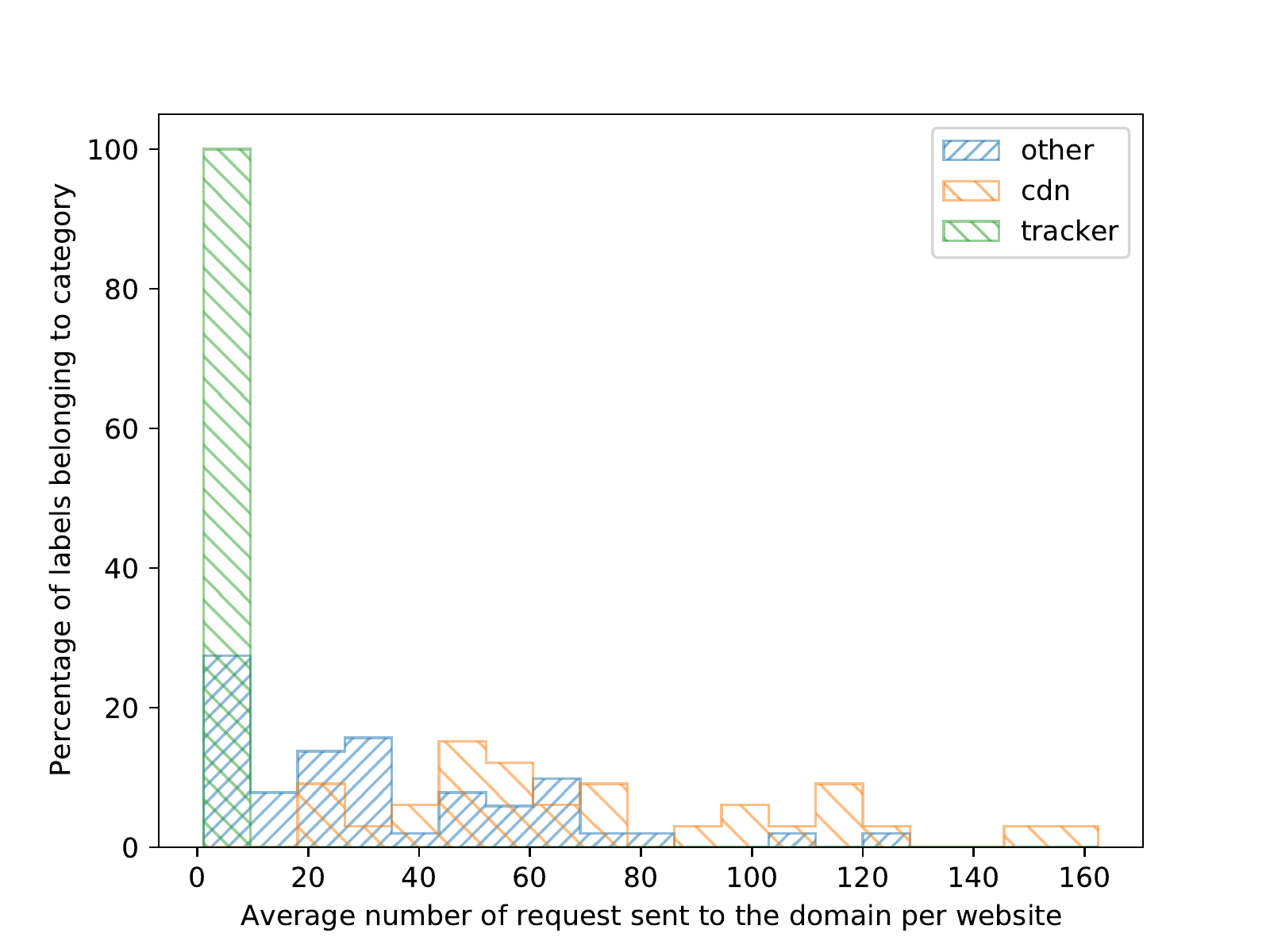}
		\caption[]
		{{\small Distribution of the average number of requests per website to the service}}    
		\label{fig:avg_req_per_page}
	\end{subfigure}
	\vskip\baselineskip
	\begin{subfigure}[b]{0.45\linewidth}   
		\includegraphics[width=\linewidth]{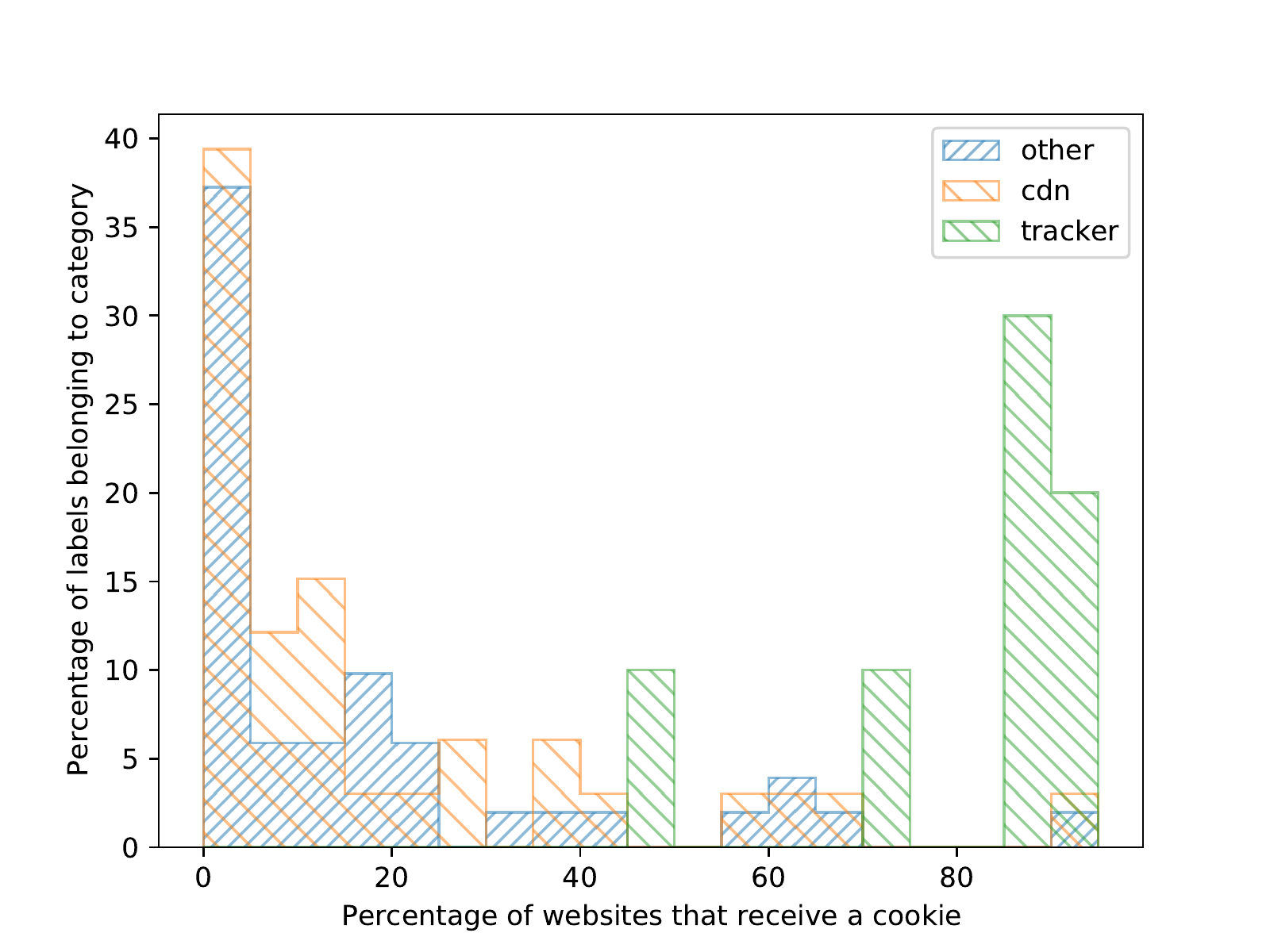}
		\caption[]
		{{\small Distribution of the percentage of responses containing at least one cookie}}    
		\label{fig:perc_pages_with_response_cookie}
	\end{subfigure}
	\qquad
	\begin{subfigure}[b]{0.45\linewidth}   
		\includegraphics[width=\linewidth]{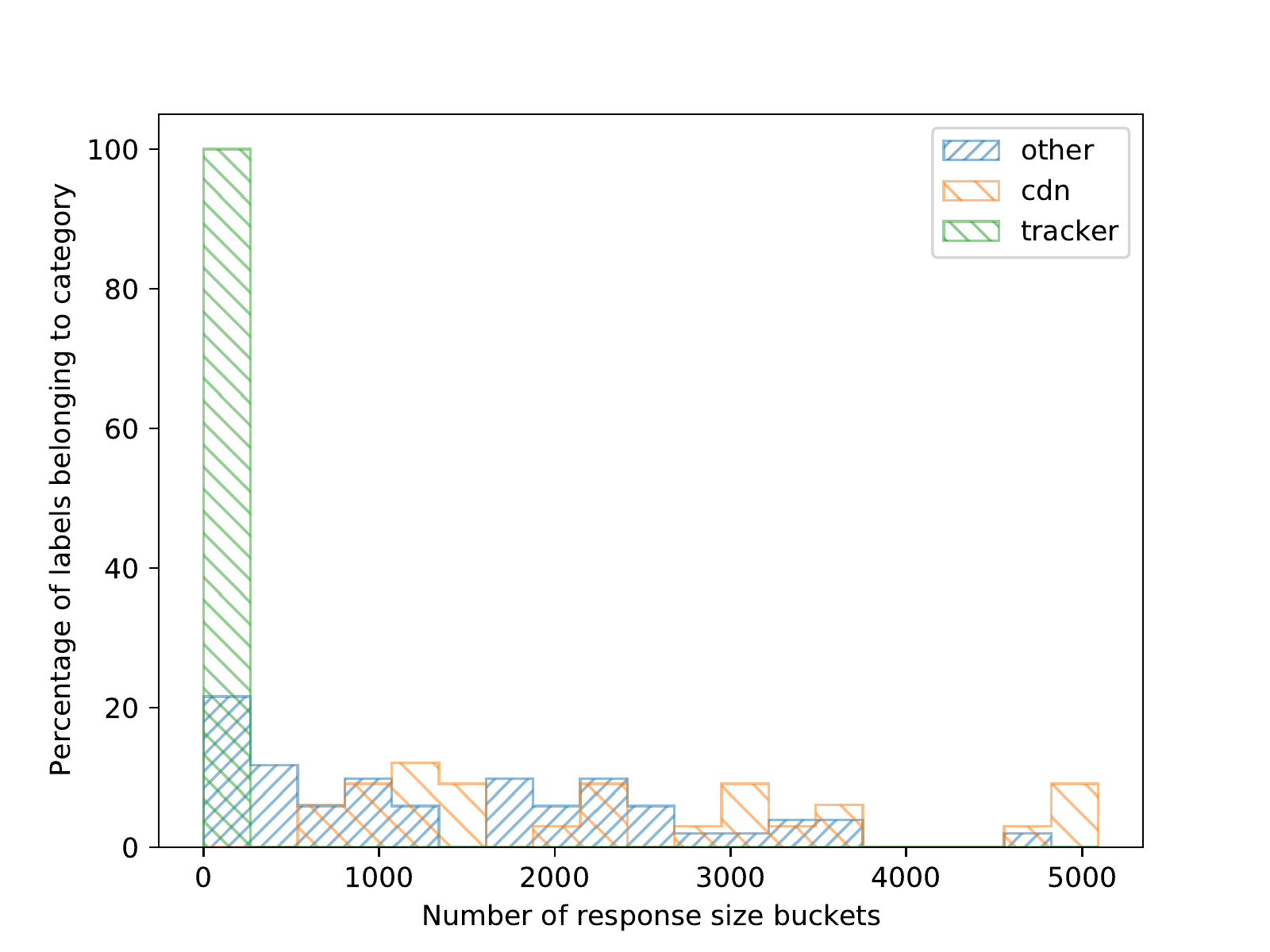}
		\caption[]
		{{\small Distribution of different response sizes sorted in buckets of size 100 bytes}}    
		\label{fig:num_resp_buckets}
	\end{subfigure}
	\caption[ The average and standard deviation of critical parameters ]
	{\small Features distinguishing trackers from other types of services} 
	\label{fig:assisted_detection}
\end{figure}

To gain a better understanding of the features we collected, we analyzed their distribution across the different categories. Figure \ref{fig:assisted_detection} shows the features that are the least overlapping for the three categories. \\
As can be deduced from Figure \ref{fig:num_resp_buckets} and Figure \ref{fig:avg_request_paths_per_page}, the number of response size buckets and the number of unique paths accessed by the website is much lower for trackers than for CDNs and other services. This was in line with our expectation that customer websites access a similar resource each time.  Furthermore, tracking services receive a low number of requests per website and often respond with a cookie. \\

Given the fact that we had a small list of confirmed trackers only, it was not feasible to build a classifier with the purpose of distinguishing tracking services from other types of services. However, our findings are still useful for performing assisted detection of tracking services. They form a simple heuristic for ruling out some companies from being trackers. With more data, the features that we gathered could likely be used for automatic detection.

\end{document}